\documentclass[aps,pra,twocolumn,superscriptaddress]{revtex4-1}
\usepackage{amsmath}
\usepackage{graphicx}
\usepackage{amsfonts}
\usepackage{color}
\usepackage{upgreek}
\usepackage{mathtools}
\usepackage{hyperref}
\usepackage{comment}
\hypersetup{
    colorlinks = true,
    linkcolor =blue,
	citecolor=blue, 
	urlcolor=blue 
}
\usepackage{array}

\usepackage{braket}
\usepackage{bm}
\usepackage[normalem]{ulem}

\begin{document}
	\title{Fundamental limitations in Lindblad descriptions of systems weakly coupled to baths}  
	\author{Devashish Tupkary}
	\email{djtupkary@uwaterloo.ca}
	\affiliation{Indian Institute of Science, Bangalore 560012, India }
	\affiliation{International Centre for Theoretical Sciences, Tata Institute of Fundamental Research, Bangalore 560089, India}
	\affiliation{Institute for Quantum Computing and Department of Physics and Astronomy, University of Waterloo, Waterloo, Ontario, Canada, N2L 3G1}	
	
	\author{Abhishek Dhar}
	\email{abhishek.dhar@icts.res.in}
	\affiliation{International Centre for Theoretical Sciences, Tata Institute of Fundamental Research, Bangalore 560089, India}

	\author{Manas Kulkarni}
	\email{manas.kulkarni@icts.res.in}
	\affiliation{International Centre for Theoretical Sciences, Tata Institute of Fundamental Research, Bangalore 560089, India}
		
	\author{Archak Purkayastha}
	\email{archak.p@tcd.ie}
	\affiliation{Trinity College Dublin, The University of Dublin, College Green, Dublin, Ireland}

\begin{abstract}
It is very common in the literature to write down a Markovian quantum master equation in Lindblad form to describe a system with multiple degrees of freedom and weakly connected to multiple thermal baths which can, in general, be at different temperatures and chemical potentials.  
However, the microscopically derived quantum master equation up to leading order in system-bath coupling is of the so-called Redfield form which is known to not preserve complete positivity in most cases.  Additional approximations to the Redfield equation are required to obtain a Lindblad form. We lay down some fundamental requirements for any further approximations to Redfield equation, which, if violated, leads to physical inconsistencies like inaccuracies in the leading order populations and coherences in the energy eigenbasis, violation of thermalization, violation of local conservation laws at the non-equilibrium steady state (NESS). We argue that one or more of these conditions will generically be violated in all the weak system-bath-coupling Lindblad descriptions existing in literature to our knowledge. As an example, we study the recently derived Universal Lindblad Equation (ULE) and use these conditions to show violation of local conservation laws due to inaccurate coherences but accurate populations in energy eigenbasis. Finally,  we exemplify our analytical results numerically in  an interacting open quantum spin system.
\end{abstract}

\maketitle

\section{Introduction}
A fundamental problem relevant across a wide range of fields including quantum optics \cite{LeHur2016}, thermodynamics \cite{quantum_thermodynamics}, chemistry \cite{quantum_chemistry}, engineering \cite{quantum_engineering} and biology \cite{quantum_biology}  is to describe a quantum system with multiple degrees of freedom connected to multiple thermal baths.  An approach very often taken is to derive an effective quantum master equation (QME) for the dynamics of the system assuming weak coupling between the system and the baths. The dynamics of the quantum system, as described by the QME, is often desired to be Markovian. It was shown by Gorini, Kossakowski, Sudarshan \cite{GKS1976}, and Lindblad \cite{lindblad1976} (GKSL) that a QME that preserves all properties of the density matrix and describes Markovian dynamics has to be of the form,
\begin{align}
\label{Lindblad_form}
& \frac{\partial \hat{\rho}}{\partial t}=i[\hat{\rho},\hat{H}_S]+\hat{\mathcal{L}}(\hat{\rho}), \nonumber \\
&\hat{\mathcal{L}}(\hat{\rho}) = i[\hat{\rho},\hat{H}_{LS}]+\sum_{\lambda=1}^{D^2-1}\gamma_{\lambda} \Big( \hat{L}_{\lambda} \hat{\rho} \hat{L}_{\lambda}^\dagger - \frac{1}{2} \{ \hat{L}_{\lambda}^\dagger \hat{L}_{\lambda}, \hat{\rho}   \} \Big),
\end{align}
which is commonly called a Lindblad equation. Here, $\hat{\rho}$ is the density matrix of the system, $\hat{H}_S$ is the system Hamiltonian, $\hat{H}_{LS}$ is the Hermitian contribution due to the presence of the baths, commonly called the Lamb shift term, $\hat{L}_{\lambda}$ are the Lindblad operators and $\gamma_{\lambda}$ are the rates, and $D$ is the dimension of the Hilbert space. Here we will confine to (effective) finite dimensional systems. Both $\hat{H}_{LS}$ and $\hat{L}_{\lambda}$ are operators in the system Hilbert space.  This equation preserves Hermiticity and trace of $\hat{\rho}$, as well as ensures non-negativity of all eigenvalues of  $\hat{\rho}$ at all times for all initial states of the system. The last condition crucially requires that $\gamma_{\lambda}\geq 0$ \cite{GKS1976,lindblad1976}. This result is one of the cornerstones of open quantum systems. Indeed, a large number of analytical~\cite{breuer_book, carmichael_book,Rivas_2012} and numerical~\cite{Weimer_2021,Plenio_1998,Dalibard1992, Mlmer1993} techniques  rely on describing experimental systems via Lindblad equations. 

The standard way to obtain a QME to leading order in system-bath coupling is via the Born-Markov approximation \cite{breuer_book, carmichael_book,Rivas_2012}. The QME so obtained, often called the Redfield equation (RE) \cite{redfield1965}, though reducible to a Lindblad-like form, is known to generically not satisfy complete positivity, i.e, not satisfy the requirement $\gamma_\lambda\geq 0$ \cite{Hartmann_2020_1,Eastham_2016,anderloni_2007,Gaspard_Nagaoka_1999,Kohen_1997,
Gnutzmann_1996,Suarez_1992}. This means that for certain initial states and at certain times, it will not give a positive semi-definite density matrix. To rectify this drawback, typically, further approximations are made to obtain a Lindblad equation either in the so called local or global forms, which we call local Lindblad (LLE) and eigenbasis Lindblad (ELE) equations respectively \cite{breuer_book, carmichael_book,Rivas_2012}. Several shortcomings of the LLE and ELE so obtained, in particular their failure to correctly describe the steady state when coupled to multiple baths, have been pointed out in  the literature, and their regimes of validity discussed \cite{Walls1970,Wichterich_2007,Rivas_2010,barranco_2014,Levy2014,archak,
Trushechkin_2016,
Eastham_2016,Hofer_2017,Gonzalez_2017,Mitchison_2018,
Cattaneo_2019,Hartmann_2020_1,Benatti_2020,konopik_2020local,
Scali_2021,Floreanini_2021,trushechkin2021,Davidovic_2021}. Despite this, the conventional wisdom is that, in principle, it should be possible to find a Lindblad equation, different from both LLE and ELE, which accurately describes the steady state. Indeed, there has been a  number of recent attempts towards developing such new variants of Lindblad equations~\cite{ule,Kleinherbers_2020,Davidovic_2020,mozgunov2020,mccauley2020,
kirvsanskas2018}, which are intended to be as accurate as the RE. 

Here, we lay down some fundamental requirements on any such attempt to recover complete positivity, which, if violated, causes physical inconsistencies like inaccuracies in the leading order populations (diagonal elements of density matrix) and coherences (off-diagonal elements of density matrix) in the energy eigenbasis, violation of thermalization and violation of local conservation laws in the non-equilibrium steady state (NESS). We show that the RE does not have any of the above physical inconsistencies, despite being generically not completely positive. On the other hand, we argue that no existing weak system-bath coupling Lindblad description, to our knowledge, generically satisfies all the conditions, and therefore has one or more of the physical inconsistencies despite being completely positive. As an example, we study in detail the so called Universal Lindblad equation (ULE) \cite{ule}, which has been recently rigorously derived to have accuracy comparable to that of the RE. We show that the ULE gives inaccurate coherences in the energy eigenbasis to the leading order and violates local conservation laws. All the above statements are shown in complete generality for time-independent Hamiltonians, without writing down any specific system Hamiltonian. This model-independent discussion sets apart our work from most previous works on checking accuracy of various QMEs,  where the accuracies were studied numerically in specific models \cite{Walls1970,Wichterich_2007,Rivas_2010,barranco_2014,Levy2014,archak,
Trushechkin_2016,
Eastham_2016,Hofer_2017,Gonzalez_2017,Mitchison_2018,
Cattaneo_2019,Hartmann_2020_1,Benatti_2020,konopik_2020local,
Scali_2021,Floreanini_2021}. We finally exemplify our discussion numerically in a three-site XXZ model coupled to two bosonic baths.

The paper is organized as follows. In Sec.\ref{sec:fundamental requirements}, we discuss the fundamental requirements. In Sec.\ref{sec: RE}, we discuss how the RE satisfies all the fundamental requirements except complete positivity. In Sec.\ref{sec:ULE}, we show that the ULE violates some of the fundamental requirements while restoring complete positivity and  make some general comments about other Lindblad equations. In Sec.\ref{sec:numerical} we numerically exemplify our discussions using the XXZ-model. In Sec.\ref{sec:summary_and_discussions}, we summarize and conclude.

\section{Fundamental requirements for an accurate weak-coupling Markovian description}
\label{sec:fundamental requirements}

\begin{figure}
\includegraphics[width=\columnwidth]{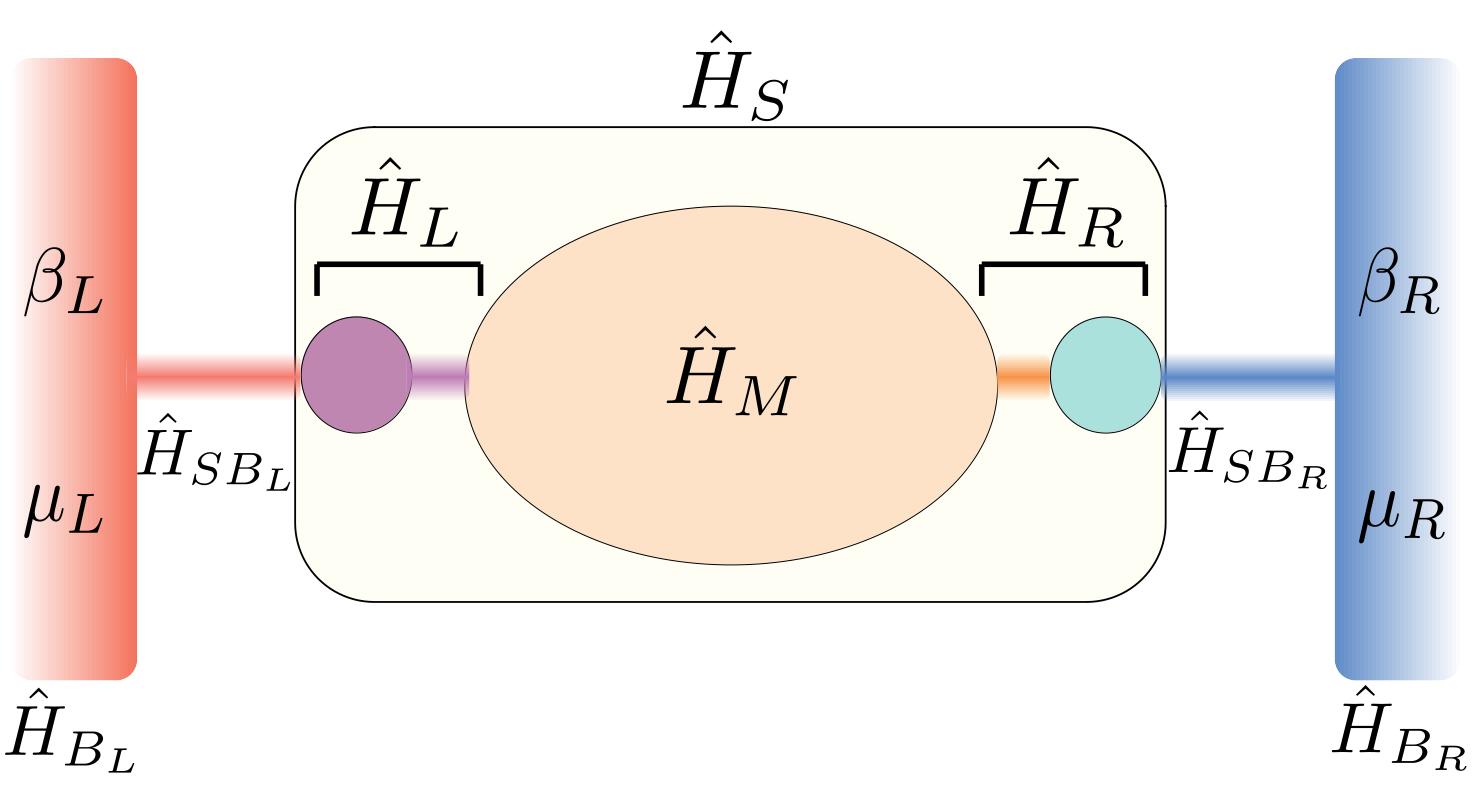}
\caption{A typical two-terminal non-equilibrium set-up with not all sites attached to the baths. The Hamiltonian $\hat{H}_M$ describes part of $\hat{H}_S$ that commutes with the system-bath coupling Hamiltonians $\hat{H}_{SB_L}$ and $\hat{H}_{SB_R}$, while the Hamiltonian $\hat{H}_L$ ($\hat{H}_R$) describes part of the system Hamiltonian that does not commute with $\hat{H}_{SB_L}$ ($\hat{H}_{SB_R}$). \label{fig:maintwo_terminal_schematic}  }
\end{figure}

\subsection{The set-up and assumptions}
For simplicity and concreteness, let us consider  a typical two-terminal set-up of the form given in Fig.~\ref{fig:maintwo_terminal_schematic}. 
The Hamiltonian of the full set-up can be written in the form
\begin{align}
& \hat{H}=\hat{H}_S + \epsilon\hat{H}_{SB} + \hat{H}_B \nonumber \\
& \hat{H}_{SB} = \hat{H}_{SB_L} + \hat{H}_{SB_R},~~\hat{H}_B = \hat{H}_{B_L} + \hat{H}_{B_R},
\end{align}
where $\hat{H}_S$ is the Hamiltonian of the system, $\hat{H}_{B_L}$ ($\hat{H}_{B_R}$) is the Hamiltonian of the left (right) bath, $\hat{H}_{SB_L}$ ($\hat{H}_{SB_R}$) is the coupling between the left (right) bath and the system, and the  dimensionless parameter $\epsilon$ controls the strength of the system-bath couplings.  We consider the system bath couplings to be weak, $\epsilon \ll 1$. The system Hamiltonian is further broken into 
\begin{align}
\hat{H}_S = \hat{H}_L  + \hat{H}_M + \hat{H}_R,
\end{align}
where $\hat{H}_M$ contains the part of the system Hamiltonian that commutes with the system-bath coupling Hamiltonians $\hat{H}_{SB_L}$ and $\hat{H}_{SB_R}$, while, $\hat{H}_L$ ($\hat{H}_R$) contains the part of the system Hamiltonian that does not commute with $\hat{H}_{SB_L}$ ($\hat{H}_{SB_R}$). We will further assume for simplicity that the system Hamiltonian $\hat{H}_S$ has no degeneracies. Initially the system is at some arbitrary state while the baths are at inverse temperatures $\beta_L$ and $\beta_R$, and chemical potentials $\mu_L$ and $\mu_R$.  We take the total number of particles (excitations) in the whole set-up to be conserved.  Without loss of generality, it is possible to assume that ${\rm Tr}\left(\hat{H}_{SB}\hat{\rho}(0)\otimes \hat{\rho}_B\right)=0$ \cite{breuer_book}, where $\hat{\rho}_B$ denotes the composite state of the two baths given by product of their individual thermal states. The state of the system at a time $t$ is 
\begin{align}
\hat{\rho}(t) = Tr_{\rm B}\left(e^{-i\hat{H}t} \hat{\rho}(0)\otimes\hat{\rho}_B e^{i\hat{H}t}\right),
\end{align}
where $Tr_{\rm B}(...)$ implies trace over bath degrees of freedom.  We will assume that in the long-time limit, the system reaches a unique NESS. This assumption physically necessitates that the system size is finite, while the baths are in the thermodynamic limit. The NESS density matrix is then defined as
\begin{align}
\label{def_NESS}
\hat{\rho}_{\rm NESS} = \lim_{t\rightarrow \infty} Tr_{\rm B}\left(e^{-i\hat{H}t} \hat{\rho}(0)\otimes\hat{\rho}_B e^{i\hat{H}t}\right).
\end{align}
 With this concrete, but still fairly general set-up and the assumptions in mind, we now look at what are the fundamental requirements for a weak-coupling-QME to accurately describe the NESS of the system.

\subsection{The fundamental requirements}
\label{subsec:fundament_requirements}

We will like to have a Markovian QME, written to leading order in system-bath coupling, that accurately predicts the NESS density matrix. To this end, we require the following physical conditions.

{\it (a) Preservation of all properties of density matrix ---} We want a QME of the form
\begin{align}
\label{Effective_QME}
\frac{\partial \hat{\rho}}{\partial t}= i[\hat{\rho}, \hat{H}_S] + \epsilon^2 \hat{\mathcal{L}}_2(\rho),
\end{align} 
where we have made it explicit that this equation is written to $O(\epsilon^2)$, which can be shown to be leading order \cite{breuer_book}. To preserve all properties of the density matrix: (i) Hermiticity, (ii) trace and (iii) complete positivity, $\hat{\mathcal{L}}_2(\rho)$ must be reducible to the form of $\hat{\mathcal{L}}(\rho)$ given in Eq.(\ref{Lindblad_form}), with $\gamma_\lambda \geq 0$.

{\it (b) Correct populations and coherences in energy eigenbasis to the leading order---} 
Since we are making a weak-coupling approximation at the level of the QME, we cannot expect to have accurate results at NESS to all orders in $\epsilon$. But, we will like to have results correct to at least the leading order. In particular, we will like to have correct results for both populations and coherences in the energy eigenbasis at least to the leading order. Coherences in the energy eigenbasis of the system, as well as population inversions in the energy eigenbasis, have been considered a resource in quantum thermodynamics and information \cite{NnC,Bennett2000,Streltsov_2017,Lostaglio_2015,Narasimhachar_2015,
Mitchison_2015,Allahverdyan_2004,Korzekwa_2016,Kammerlander_2016,
Santos_2019,Francica_2019}. Thus, it is important to describe populations and coherences correctly. Moreover, as we will see later, both populations and coherences are crucially linked to having correct currents at NESS.

{\it (c) Thermalization ---}
 Further, having correct populations to leading order is also linked with the fundamental phenomenon of thermalization of the system with the baths when they have same temperatures and chemical potentials $\beta_L=\beta_R=\beta$, $\mu_L=\mu_R=\mu$. In that case, on physical grounds, the steady state density matrix is expected to satisfy the following condition
\begin{align}
\label{Thermalization_condition}
\lim_{\epsilon\rightarrow 0} \Big(\lim_{t\rightarrow \infty} \hat{\rho}(t)\Big)~=~\frac{e^{-\beta(\hat{H}_S-\mu \hat{N}_S)}}{{\rm Tr}\left( e^{-\beta(\hat{H}_S-\mu \hat{N}_S)} \right)}.
\end{align}
where $\hat{N}_S$ is the total particle or excitation number operator of the system. In other words,  populations of the density matrix in the energy eigenbasis of the system, to the leading order in system-bath coupling, should follow a Gibbs distribution specified by the temperature and the chemical potential of the baths when they are all equal. As we will show later, if the steady state is unique, this condition is always satisfied in the exact steady state (Eq.\ref{def_NESS}). Note that, in Eq.(\ref{Thermalization_condition}), the order of limits cannot be interchanged.

%


{\it (d) Preservation of local conservation laws ---} Writing the Heisenberg's equation of motion for our set-up and taking trace,  it is clear that the rate of change of expectation value of any operator $\hat{O}$ which commutes with the system-bath couplings is given by
\begin{align}
\label{Local_conservation_condition}
\frac{d \langle \hat{O} \rangle}{dt} = -i\langle[\hat{O}, \hat{H}_S]\rangle,~~\forall~~[\hat{O},\hat{H}_{SB}]=0,
\end{align}
where $\langle \ldots \rangle={\rm Tr}(\hat{\rho} \ldots)$. 
This condition is obviously true irrespective of strength of system-bath couplings. We will like our effective weak system-bath coupling description to preserve this condition. Combining Eq.(\ref{Local_conservation_condition}) and Eq.(\ref{Effective_QME}), we see that this entails
\begin{align}
\label{Local_conservation_condition_on_L}
{\rm Tr}\left( \hat{O} \hat{\mathcal{L}_2}(\hat{\rho}) \right) = 0,~~\forall~~[\hat{O},\hat{H}_{SB}]=0.
\end{align}
This condition is not usually discussed in the context of deriving weak-coupling Markovian descriptions. However, the importance of this condition becomes clear when it is written for one of the locally conserved quantities, such as $\hat{H}_M$,
\begin{align}
\label{current_defs}
& \frac{d \langle \hat{H}_M \rangle}{dt} =- i\langle[\hat{H}_M, \hat{H}_S]\rangle = J_{L\rightarrow M} - J_{M \rightarrow R}, \nonumber \\
& J_{L\rightarrow M}=  -i\langle[\hat{H}_M, \hat{H}_L]\rangle,~~ J_{M \rightarrow R}= i\langle[\hat{H}_M, \hat{H}_R] \rangle,
\end{align}
where we have defined the energy current into the central region from the left as $J_{L\rightarrow M}$, and that from the central region to the right as $J_{M\rightarrow R}$. These currents are expectation values of some system operators and do not involve any explicit dependence on system-bath couplings.  At NESS, the rate of change of expectation values of all system operators is zero. Thus, $\frac{d \langle \hat{H}_M \rangle}{dt}=0$, which implies $J_{L\rightarrow M}=J_{M \rightarrow R}$. So we have,
\begin{align}
\label{internal_NESS_current_condition}
\langle[\hat{H}_M, \hat{H}_L]\rangle = -\langle[\hat{H}_M, \hat{H}_R]\rangle\neq 0.
\end{align}
This is a fundamental property of the NESS that should exactly hold irrespective of strength of system bath couplings. The Eq.(\ref{current_defs}), which at NESS leads to the above condition, is a continuity equation coming from local conservation of energy. Analogous conditions can be derived for other locally conserved quantities. The QME written to leading order should satisfy all such conditions. It is clear that Eq.(\ref{Local_conservation_condition_on_L}) guarantees this.

It can be seen that satisfying Eq.(\ref{Local_conservation_condition_on_L}), and giving correct rate of change of expectation value of $\hat{O}$ in Eq.(\ref{Local_conservation_condition}) requires that the coherences in the energy eigenbasis are given correctly. To see this, we write the system Hamiltonian in the energy eigenbasis, 
\begin{align}
\hat{H}_S = \sum_{\alpha} E_\alpha |E_\alpha \rangle \langle E_\alpha |.
\end{align}
Using this basis, Eq.(\ref{Local_conservation_condition}) can be written as
\begin{align}
\label{rate_and_coherence}
& \frac{d \langle \hat{O} \rangle}{dt} = -i \sum_{\alpha,\nu=1}^D (E_\alpha - E_\nu) \langle E_\alpha | \hat{\rho} | E_\nu \rangle \langle E_\nu | \hat{O} | E_\alpha \rangle ,\nonumber \\
&\forall~~[\hat{O},\hat{H}_{SB}]=0.
\end{align}
Clearly, the right-hand-side of above equation is governed by the coherences between in the energy eigenbasis of the system. Thus, satisfying local conservation laws and obtaining correct currents at NESS requires the coherences to be correct (at least to leading order).

\begin{table*}
\begin{tabular}{|c|c|m{10.7cm}|} 
\hline
1. &  Correct	populations to leading order [$O(\epsilon^0)$] & \qquad $\left\langle E_{\alpha} \left| \hat{\mathcal{L}}_2^\prime\left[\sum_\alpha p_\alpha |E_\alpha\rangle \langle E_\alpha |\right] \right| E_{\alpha} \right\rangle=\left\langle E_{\alpha} \left| \hat{\mathcal{L}}_2\left[\sum_\alpha p_\alpha |E_\alpha\rangle \langle E_\alpha |\right] \right| E_{\alpha} \right\rangle$,\\
\hline
2. & Correct coherences to leading order [$O(\epsilon^2)$] & 	$\left\langle E_{\alpha} \left| \hat{\mathcal{L}}_2^\prime\left[\sum_\alpha p_\alpha |E_\alpha\rangle \langle E_\alpha |\right] \right| E_{\nu} \right\rangle=\left\langle E_{\alpha} \left| \hat{\mathcal{L}}_2\left[\sum_\alpha p_\alpha |E_\alpha\rangle \langle E_\alpha |\right] \right| E_{\nu} \right\rangle~\forall~\alpha \neq \nu$. \\
\hline
3. & Thermalization [Eq.\eqref{Thermalization_condition})] & \hspace*{40pt} $\left\langle E_{\alpha} \left| \hat{\mathcal{L}}_2^\prime\left[\sum_\alpha p_\alpha |E_\alpha\rangle \langle E_\alpha |\right] \right| E_{\alpha} \right\rangle=0 \Rightarrow p_\alpha \propto e^{-\beta(E_\alpha - \mu N_\alpha)}$ \newline \hspace*{80pt} when $\beta_L=\beta_R=\beta$, $\mu_L=\mu_R=\mu$. \\
\hline
4. & Preservation of local conservation laws & \hspace*{80pt} ${\rm Tr}\left( \hat{O} \hat{\mathcal{L}}_2^\prime(\hat{\rho}) \right) = 0,~~\forall~~[\hat{O},\hat{H}_{SB}]=0.$\\
\hline
5. & Complete positivity & \hspace*{40pt} $\hat{\mathcal{L}}_2^\prime[\hat{\rho}] = i[\hat{\rho},\hat{H}_{LS}]+\sum_{\lambda=1}^{D^2-1}\gamma_{\lambda} \Big( \hat{L}_{\lambda} \hat{\rho} \hat{L}_{\lambda}^\dagger - \frac{1}{2} \{ \hat{L}_{\lambda}^\dagger \hat{L}_{\lambda}, \hat{\rho}   \} \Big)$, $\gamma_\lambda \geq 0$. \\
\hline
\end{tabular}
\caption{Summary of necessary conditions that a QME written to the leading order in system-bath coupling must satisfy to accurately describe the steady state of a set-up of the form in Fig.\ref{fig:maintwo_terminal_schematic}. In above, $\hat{\mathcal{L}}_2$ is the $O(\epsilon^2)$ term obtained by systematically doing time-convolution-less expansion and taking long-time limit [see Eqs.\eqref{general_TCL_QME}, \eqref{exact_NESS_equation}]. The $\hat{\mathcal{L}}_2[\hat{\rho}]$ coincides with that obtained in Redfield equation via Born-Markov approximation and is known to generically not satisfy the complete positivity requirement $\gamma_\lambda\geq 0$. The $\hat{\mathcal{L}}_2^\prime$ is any modification of $\hat{\mathcal{L}}_2$ which may restore complete positivity. \label{table1}}  
\end{table*}

\subsection{The equations giving correct populations and coherences to the leading order}
Let us now see what condition in terms of $\hat{\mathcal{L}}_2$ must be satisfied so that the populations are the coherences in the energy eigenbasis are correct to the leading order. We assume that the NESS density matrix defined in Eq.(\ref{def_NESS}) and can be  expanded in powers of $\epsilon$, 
\begin{equation}
\label{eq:rho_ness_general}
\hat{\rho}_{\rm NESS} = \sum_{m=0}^{\infty} \epsilon^{2m} \hat{\rho}_{\rm NESS}^{(2m)}.
\end{equation}
The QME  describing the evolution of $\hat{\rho}(t)$ can also be systematically expanded in the so-called time-convolution-less form \cite{breuer_book}, 
\begin{equation}
	\frac{\partial\hat{\rho}}{\partial t}= \sum_{m=0}^{\infty} \epsilon^{2m} \hat{{\mathcal{L}}}_{2m}(t)[\hat{\rho}(t)], \label{general_TCL_QME}
\end{equation} where $\hat{{\mathcal{L}}}_{2m}(t)$ are in general time dependent linear operators and 
$\hat{{\mathcal{L}}}_{0}(t)[\hat{\rho}(t)]=i[\hat{\rho}(t),\hat{H}_S]$. By definition, the steady state $\hat{{\rho}}_{\rm NESS}$ satisfies
\begin{align}
	\label{exact_NESS_equation}
	0=\sum_{m=0}^{\infty} \epsilon^{2m} {\hat{\mathcal{L}}}_{2m}[\hat{\rho}_{\rm NESS}],
\end{align}
where we denote $\hat{\mathcal{L}}_{2m}\equiv \hat{{\mathcal{L}}}_{2m}(t\to\infty)$.

Using the  series expansion of $\hat{{\rho}}_{\rm NESS}$ and matching terms order-by-order yields, at the lowest order ($\epsilon^0$), the condition 
$[\hat{{\rho}}_{\rm NESS}^{(0)}, \hat{H}_S]=0.$  This implies that $\hat{{\rho}}_{\rm NESS}^{(0)}$ is diagonal in the energy eigenbasis of the system, 
\begin{align}
\langle E_\alpha |\hat{{\rho}}_{\rm NESS}^{(0)} | E_{\nu} \rangle = 0~~\forall~\alpha \neq \nu.
\end{align}
This shows that while the populations can have $O(\epsilon^0)$ term, the leading order contribution to coherences in energy eigenbasis is $O(\epsilon^2)$. (Remember that we have assumed no degeneracies in the system Hamiltonian.) This means $\hat{{\rho}}_{\rm NESS}^{(0)}$ can be written as 
\begin{align}
\label{rho_NESS_0}
\hat{\rho}_{\rm NESS}^{(0)}=\sum_{\alpha} p_\alpha |E_\alpha\rangle \langle E_\alpha |.
\end{align}
As a consequence, when the temperatures and the chemical potentials of the baths are equal, $\beta_L=\beta_R=\beta$, $\mu_L=\mu_R=\mu$, the thermalization condition Eq.\eqref{Thermalization_condition} requires,
\begin{align}
p_\alpha \propto e^{-\beta(E_\alpha - \mu N_\alpha)},
\end{align}
where $N_\alpha$ is the number of particles or excitations in the energy eigenstate $E_\alpha$.

Let us now look at two higher orders in $\epsilon$. At  order $\epsilon^2$, using Eq.(\ref{eq:rho_ness_general}) in Eq.~(\ref{exact_NESS_equation}) then leads to the conditions: 
 \begin{align}
 	& \left\langle E_{\alpha} \left| \hat{\mathcal{L}}_2[\hat{{\rho}}_{\rm NESS}^{(0)}] \right| E_{\alpha} \right\rangle=0, \label{exact_1} \\
 	& i(E_\alpha -E_\nu) \left\langle E_{\alpha} \left| \hat{{\rho}}_{\rm NESS}^{(2)}\right| E_{\nu} \right\rangle  \nonumber \\ & +\left\langle E_{\alpha} \left| \hat{\mathcal{L}}_2[\hat{{\rho}}_{\rm NESS}^{(0)}] \right| E_{\nu} \right\rangle=0 \quad \forall~~\alpha \neq \nu, \label{exact_2}
\end{align}
and at order $\epsilon^4$ we get:
\begin{align}
 	& \left\langle E_{\alpha} \left| \hat{\mathcal{L}}_2[\hat{{\rho}}_{\rm NESS}^{(2)}] \right| E_{\alpha} \right\rangle+\left\langle E_{\alpha} \left| \hat{\mathcal{L}}_4[\hat{{\rho}}_{\rm NESS}^{(0)}] \right| E_{\alpha} \right\rangle=0. \label{exact_3}
 \end{align}
The set of equations obtained by writing Eq.~\eqref{exact_1} for various values of eigenstate index $\alpha$, gives a complete set of equations which determines the non-zero elements of $\hat{\rho}_{\rm NESS}^{(0)}$. Given $\hat{\rho}_{\rm NESS}^{(0)}$, the set of equations obtained by writing Eq.~\eqref{exact_2} for various values of eigenstate indices $\alpha$ and $\nu$,  determines the off-diagonal elements of $\hat{\rho}_{\rm NESS}^{(2)}$ in the energy eigenbasis. Finally, given the off-diagonal elements of $\hat{\rho}_{\rm NESS}^{(2)}$ in the energy eigenbasis, and the $\hat{\rho}_{\rm NESS}^{(0)}$,  the set of equations obtained by writing Eq.~\eqref{exact_3} for various values of the eigenstate index $\alpha$ determines the diagonal elements of $\hat{\rho}_{\rm NESS}^{(2)}$. It is crucial to note the occurrence of $\hat{\mathcal{L}}_4$ in Eq.~\eqref{exact_3}. Since the leading order populations are $O(\epsilon^0)$ and the leading order coherences are $O(\epsilon^2)$, Eqs.\eqref{exact_1} and \eqref{exact_3} give the necessary equations to solve in order to obtain populations and coherences correct to the leading order. The above discussion correspond to the NESS obtained from the exact QME, Eq.~\eqref{general_TCL_QME}, in the regime of small $\epsilon$.

\subsection{The need to go beyond standard Lindblad descriptions}


In the previous subsection, $\hat{\mathcal{L}}_2$ was derived from a time-convolution-less expansion. This fixes the form of $\hat{\mathcal{L}}_2$. If  we truncate the time-convolution-less expansion at $O(\epsilon^2)$ and take the long-time limit, we get an equation exactly of our desired form in Eq.\eqref{Effective_QME}. This coincides with Born-Markov approximation and gives the RE \cite{breuer_book}. However, it is known that the RE does not generically satisfy the requirement of complete positvity \cite{Hartmann_2020_1,Eastham_2016,anderloni_2007,Gaspard_Nagaoka_1999,Kohen_1997,
Gnutzmann_1996,Suarez_1992}. Due to this, very often further approximations are made on $\hat{\mathcal{L}}_2$, modifying it to, say, $\hat{\mathcal{L}}_2^\prime$  so as to impose complete positivity. Our analysis in previous subsections allows us to impose some necessary restrictions on the approximations by fixing certain components of  $\hat{\mathcal{L}}_2^\prime$, so that the fundamental requirements are satisfied. The set of conditions is summarized in Table~\ref{table1}. In particular, we see that for both populations and coherences to be correct to the leading order, we require the operation of  $\hat{\mathcal{L}}_2^\prime$ on any state diagonal in the energy eigenbasis to be same as that of $\hat{\mathcal{L}}_2$. 

In most studies of accuracy of results from QMEs, the accuracy is checked by applying them to some particular system and comparing results with some other more accurate method \cite{Walls1970,Wichterich_2007,Rivas_2010,barranco_2014,Levy2014,archak,
Trushechkin_2016,
Eastham_2016,Hofer_2017,Gonzalez_2017,Mitchison_2018,
Cattaneo_2019,Hartmann_2020_1,Benatti_2020,konopik_2020local,
Scali_2021,Floreanini_2021,trushechkin2021}. This does not let one clearly comment on situations where such more accurate methods are unavailable. The set of conditions in Table~\ref{table1} allows us to assess the accuracy of $\hat{\mathcal{L}}_2^\prime$ by making formal checks, without explicitly writing it down for any particular system. They therefore allow a completely general discussion of accuracy of NESS results from QMEs, even in cases where some more accurate method is unavailable.

The most standard approximations on $\hat{\mathcal{L}}_2$ involve reducing it to either the form of LLE or ELE. As mentioned before, it is already well-established that the LLE and the ELE do not satisfy all the fundamental requirements mentioned in Sec.~\ref{subsec:fundament_requirements} \cite{Walls1970,Wichterich_2007,Rivas_2010,barranco_2014,Levy2014,archak,
Trushechkin_2016,
Eastham_2016,Hofer_2017,Gonzalez_2017,Mitchison_2018,
Cattaneo_2019,Hartmann_2020_1,Benatti_2020,konopik_2020local,
Scali_2021,Floreanini_2021,trushechkin2021}, and therefore do not satisfy all the conditions in Table~\ref{table1}. The LLE does not show thermalization, while it explicitly satisfies the fourth condition in Table~\ref{table1}, thereby preserving local conservation laws. On the other hand, the ELE explicitly neglects various coherences in the energy eigenbasis of the system, but satisfies thermalization \cite{breuer_book}. Moreover, ELE does not also satisfy the fourth condition in Table~\ref{table1}. Further, the ELE is known to not satisfy Eq.\eqref{internal_NESS_current_condition}, because it gives zero currents inside the system even in an out-of-equilibrium condition \cite{Wichterich_2007,archak}.  So neither of these standard Lindblad equations satisfy all the requirements above. This makes it necessary to go beyond these standard approximations, even at weak system-bath coupling.

In the following section, we consider the RE in more detail and show that even though the RE does not generically satisfy the requirement of complete positivity, it satisfies all other of the fundamental requirements.

\section{The Redfield equation}
\label{sec: RE}

\subsection{Accuracy of populations and coherences from the Redfield equation and lack of complete positivity}
\label{second_order_accuracy}

As mentioned before, the RE is obtained by  truncating Eq.(\ref{general_TCL_QME}) at second order and replacing $\hat{{\mathcal{L}}}_{2}(t)$ by  $\hat{{\mathcal{L}}}_{2}(t\to\infty) \equiv \hat{{\mathcal{L}}}_{2}$  \cite{breuer_book},  
 \begin{align}
 	\label{TCL2}
 	\frac{\partial{\hat{{\rho}}}}{\partial t}= \hat{\mathcal{L}}_{0}[\hat{{\rho}}(t)]+ \epsilon^2 \hat{\mathcal{L}}_{2}[\hat{{\rho}}(t)].
 \end{align}
The NESS obtained from the RE, which we denote by $\hat{\widetilde{\rho}}_{\rm NESS}$, is given by
\begin{align}
\hat{\widetilde{\rho}}_{\rm NESS} = \lim_{t\rightarrow \infty} e^{t (\hat{\mathcal{L}}_0 + \hat{\mathcal{L}}_2)} [{\rho}(0)], 
\end{align}
and satisfies
\begin{align}
 	\label{TCL2_NESS_equation}
 	0=\hat{\mathcal{L}}_{0}[\hat{\widetilde{{\rho}}}_{\rm NESS}]+ \epsilon^2 
\hat{\mathcal{L}}_{2} [\hat{\widetilde{{\rho}}}_{\rm NESS}].
 \end{align}
The density matrix can also be expanded in powers of $\epsilon$, $\hat{\widetilde{\rho}}_{\rm NESS} = \sum_{m=0}^{\infty} \epsilon^{2m} \hat{\widetilde{\rho}}_{\rm NESS}^{(2m)}$.
The question then is, to what order in $\epsilon$ does elements of $\hat{\widetilde{\rho}}_{\rm NESS}$ and $\hat{{\rho}}_{\rm NESS}$ agree.  

Proceeding as before we again get $[\hat{\widetilde{\rho}}^{(0)}_{\rm NESS}, \hat{H}_S]=0$ implying a diagonal $\hat{\widetilde{\rho}}^{(0)}_{\rm NESS}$ in energy eigenbasis. At $O(\epsilon^2)$  we  now obtain:
 \begin{align}
 	& \left\langle E_{\alpha} \left| {\hat{\mathcal{L}}}_2[\hat{{\widetilde{\rho}}}_{\rm NESS}^{(0)}] \right| E_{\alpha} \right\rangle=0, \label{secondorder_1} \\
 	& i(E_\alpha -E_\nu) \left\langle E_{\alpha} \left| \hat{{\widetilde{\rho}}}_{\rm NESS}^{(2)}\right| E_{\nu} \right\rangle \nonumber \\&+\left\langle E_{\alpha} \left| \hat{\mathcal{L}}_2[\hat{{\widetilde{\rho}}}_{\rm NESS}^{(0)}] \right| E_{\nu} \right\rangle=0\quad  \forall~~\alpha \neq \nu, \label{secondorder_2}
 \end{align} 
while at $O(\epsilon^4)$ we get
\begin{align}
 	& \left\langle E_{\alpha} \left| \hat{\mathcal{L}}_2[\hat{{\widetilde{\rho}}}_{\rm NESS}^{(2)}] \right| E_{\alpha} \right\rangle =0. \label{secondorder_3}
 \end{align} 
Equations~\eqref{secondorder_1}, \eqref{secondorder_2} and \eqref{secondorder_3} are the analogs of Eqs.~(\ref{exact_1}), (\ref{exact_2}), and (\ref{exact_3}) respectively.  We see from Eqs.(\ref{exact_1}) and (\ref{secondorder_1}) that $\hat{{\rho}}^{(0)}_{\rm NESS}$ and $\hat{\tilde\rho}^{(0)}_{\rm NESS}$  satisfies the exact same set of equations, which implies 
 \begin{align}
 	\hat{\widetilde\rho}_{\rm NESS}^{(0)}=\hat{\rho}_{\rm NESS}^{(0)}.
 \end{align}
 Next, from Eqs.(\ref{exact_2}) and (\ref{secondorder_2}), we see that the off-diagonal elements of $\hat{\rho}_{\rm NESS}^{(2)}$ and $\hat{\widetilde\rho}_{\rm NESS}^{(2)}$ in energy eigenbasis also satisfy exactly same equation therefore implying,
 \begin{align}
 	\label{eq:coherences}
 	\left\langle E_{\alpha} \left| \hat{\widetilde\rho}_{\rm NESS}^{(2)}\right| E_{\nu} \right \rangle=\left\langle E_{\alpha} \left| \hat{\rho}_{\rm NESS}^{(2)}\right| E_{\nu} \right\rangle ,~~\forall~~\alpha \neq \nu.
 \end{align} 
 Thus, the leading order terms in populations and coherences are given correctly by the RE.
 However, Eq.(\ref{exact_3}) which fixes the diagonal elements of $\hat{\rho}_{\rm NESS}^{(2)}$ in energy eigenbasis is different from Eq.(\ref{secondorder_3}) which fixes the same for $\hat{\widetilde\rho}_{\rm NESS}^{(2)}$. So, 
 \begin{align}
 	\left\langle E_{\alpha} \left| \hat{{{\widetilde\rho}}}_{\rm NESS}^{(2)}\right| E_{\alpha} \right\rangle\neq \left\langle E_{\alpha} \left| \hat{\rho}_{\rm NESS}^{(2)}\right| E_{\alpha} \right \rangle.
 \end{align}
 Specifically, Eq.(\ref{exact_3}) shows that, to obtain the diagonal elements of NESS density matrix in the energy eigenbasis correct to $O(\epsilon^2)$, one needs the QME up of $O(\epsilon^4)$. Based on these results, we have,
 \begin{align}
 	\label{error_order}
 	|| \hat{\rho}_{\rm NESS}-\hat{\widetilde\rho}_{\rm NESS} || \sim O(\epsilon^2),
 \end{align}
 where $||\hat{P}||$ is norm of the operator $\hat{P}$. So, the error in obtaining the NESS from Eq.(\ref{TCL2}) is $O(\epsilon^2)$. Despite this, the leading order coherences in the energy eigenbasis, which are $O(\epsilon^2)$, are given correctly. Having $O(\epsilon^2)$ term in the coherences accurately while not having the corresponding correction in the populations can cause at $O(\epsilon^2)$ violation of positivity of the density matrix \cite{fleming_cummings_accuracy,Archak_2020}. Thus, the lack of complete positivity of the RE stems from  the above mismatch in order of accuracy between populations and coherences. 
 

\subsection{Thermalization}
\label{sec_thermalization}
The explicit form of Eq.(\ref{TCL2}) is given by (Eq.(9.52) of Ref.\cite{breuer_book}),
\begin{align}
	\label{RE1}
	\frac{\partial\hat{\rho}}{\partial t}=&i[ \hat{\rho}(t),{H}_S] \nonumber\\
	+& \epsilon^2 \int_0^\infty dt^\prime [\hat{H}_{SB},[\hat{H}_{SB}(-t^\prime),\hat{\rho}(t) \otimes \hat{\rho}_B]],
\end{align}
where $\hat{H}_{SB}(t)=e^{i(\hat{H}_S+\hat{H}_B)t}\hat{H}_{SB}e^{-i(\hat{H}_S+\hat{H}_B)t}$. Let us now consider the following canonical model of thermal baths: 
\begin{align}
\label{bathH}
	\hat{H}_B&= \sum_{\ell}{^{^\prime}} \sum_{r=1}^\infty \Omega^\ell_r \hat{B}^{{(\ell)} \dagger}_r \hat{B}^{(\ell)}_r,  \\
	\hat{H}_{SB} &= \sum_{\ell}{^{^\prime}} \sum_{r=1}^{\infty}  (\kappa_{\ell r} \hat{B}^{ (\ell) \dagger}_r \hat{S}_\ell + \kappa_{\ell r}^* \hat{S}_{\ell}^{ \dagger} \hat{B}^{(\ell)}_r ) \label{couplingH}
\end{align}
 where $\sum_{\ell}^\prime$  indicates sum over all sites of the system where the baths are attached, $\hat{B}^{(\ell)}_r$ is bosonic or fermionic annihilation operator for the $r$-th mode of the bath attached at the $\ell$-th site and $\hat{S}_\ell$ is the system operator coupling to the bath at site $\ell$. At initial time, the baths are taken to be in their respective thermal state with inverse temperatures $\beta_\ell$ and chemical potentials $\mu_\ell$.  The dynamics of the system can be shown to be governed by the bath spectral functions, defined as
\begin{align}
\mathfrak{J}_\ell(\omega)=  \sum_k 2 \pi \left| \kappa_{\ell k} \right|^2 \delta (\omega-\Omega^\ell_k) \label{Jw}
\end{align}
and the Fermi or Bose distributions, $n_\ell(\omega)=[e^{\beta_\ell(\omega-\mu_\ell)}\pm 1]^{-1}$,  corresponding to the initial states of the baths. The RE for this set-up is obtained by simplification of Eq.~(\ref{RE1}), using Eqs.~\eqref{bathH},\eqref{couplingH}:   
\begin{align}
	\label{RE2}
	\frac{\partial\hat{\rho}}{\partial t} &= i[\hat{\rho},\hat{H}_S] \nonumber  \\
	&- \epsilon^2 \sum_\ell{^{^\prime}} \Big ( \big[\hat{S}_\ell^\dagger, \hat{S}_\ell^{(1)} \hat{\rho}\big] + \big[\hat{\rho} \hat{S}_\ell^{(2)}, \hat{S}_\ell^\dagger \big] + {\rm h.c.} \Big),
\end{align}
where 
\begin{align}
	&\hat{S}_\ell^{(1)}=\int_0^\infty dt^\prime \int\frac{d\omega}{2\pi}\hat{S}_\ell(-t^\prime) e^{\beta_\ell(\omega-\mu_\ell)}\mathfrak{J}_\ell(\omega)n_\ell(\omega)e^{-i\omega t^\prime}, \nonumber \\  
	&\hat{S}_\ell^{(2)}=\int_0^\infty dt^\prime \int\frac{d\omega}{2\pi}\hat{S}_\ell(-t^\prime)  \mathfrak{J}_\ell(\omega)n_\ell(\omega)e^{-i\omega t^\prime}, 
\end{align}
with $\hat{S}_\ell(t)=e^{i\hat{H}_S t} \hat{S}_\ell e^{-i\hat{H}_S t}$, and h.c. denoting Hermitian conjugate. Note that, in general, $[\hat{S}_\ell,\hat{S}_\ell^{(1)}]\neq 0$, $[\hat{S}_\ell,\hat{S}_\ell^{(2)}]\neq 0$.
Let us now check if the RE satisfies the thermalization condition given in Eq.\eqref{Thermalization_condition}. To this end,  we need to find the leading order populations when the temperatures and chemical potentials of the baths are the same, $\beta_L=\beta_R=\beta$, $\mu_L=\mu_R=\mu$. Defining $E_{\nu \alpha}=E_\nu-E_\alpha$ and  writing $\hat{\rho}_{\rm NESS}^{(0)}=\sum_{\alpha} p_\alpha |E_\alpha\rangle \langle E_\alpha |$, it can be checked after some algebra that Eq.~(\ref{secondorder_1}), for the RE in Eq.~\eqref{RE2}, takes the form: 
\begin{align}
	& \sum_{\nu} \Big[|\langle E_\nu | \hat{S}_\ell |E_\alpha \rangle|^2\left[p_\alpha-e^{\beta(E_{\nu\alpha}-\mu)} p_\nu\right] \mathfrak{J}_\ell(E_{\nu\alpha})n_\ell(E_{\nu\alpha}) \nonumber \\
	& +|\langle E_\alpha | \hat{S}_\ell |E_\nu \rangle|^2\left[p_\nu-e^{\beta(E_{\alpha\nu}-\mu)}p_\alpha\right] \mathfrak{J}_\ell(E_{\alpha\nu})n_\ell(E_{\alpha\nu})\Big] \nonumber \\&=0.
\end{align}
 We now consider the case where the system operator coupling to the bath can create or annihilate a single particle, i.e $\langle E_\nu | \hat{S}_\ell |E_\alpha \rangle=0$ $\forall$ $N_\alpha-N_\nu\neq 1$. This condition ensures that with the system-bath coupling of the form in Eq.\eqref{couplingH}, the total number of exictations in the full set-up is conserved. It can then be checked that the following choice of $p_\alpha$ satisfies the above equation,
\begin{align}
	p_\alpha \propto e^{-\beta(E_\alpha - \mu N_\alpha)}.
\end{align}
Since we have assumed the steady state to be unique, this proves that the thermalization condition, i.e, the third condition in Table.~\ref{table1}, is satisfied. 

Note that, since the leading order populations obtained from RE are exactly the same as those obtained from the full general time-convolutionless-equation Eq.\eqref{general_TCL_QME}, the above proof of thermalization is essentially a proof of thermalization for the full time-convolutionless-equation, Eq.\eqref{general_TCL_QME}. Thus, if the steady state is unique, the thermalization condition is satisfied in complete generality. So, any Lindblad equation that does not satisfy this condition (such as the LLE) cannot describe the steady state of a system coupled to thermal baths. 


\subsection{Satisfying local conservation laws} 
\label{local_laws}

 Given the RE, Eq.\eqref{RE2} one can write down an expression for expectation value of any system operator ${O}$, 
\begin{align}
	\label{operator_RE}
	\frac{d\braket{\hat{O}}}{dt}=&-i\braket{[\hat{O},\hat{H}_S]}\nonumber \\
	-\epsilon^2 & \sum_\ell^\prime\Big( \left\langle[\hat{O},\hat{S}_\ell^\dagger]\hat{S}_\ell^{(1)}\right\rangle-\left\langle\hat{S}_\ell^{(2)}[\hat{O},\hat{S}_\ell^\dagger]\right\rangle \nonumber \\
	&+ \left\langle \hat{S}_\ell^{(1)^\dagger}[\hat{S}_\ell,\hat{O}]\right\rangle-\left\langle [\hat{S}_\ell, \hat{O}] \hat{S}_\ell^{(2)^\dagger}\right\rangle  \Big).
\end{align}
It is clear from above expression that the fourth condition in Table.~\ref{table1} is satisfied. Thus, it is clear that the local conservation laws inside the system are satisfied. Not only that, as we show below, further conservation laws relating currents from the baths and currents inside the system are also satisfied.

For the kind of set-up in Fig.\ref{fig:maintwo_terminal_schematic}, the RE can be written as
\begin{align}
	\label{TCL3}
	\frac{\partial\hat{\rho}}{\partial t}= i[\hat{\rho},\hat{H}_S]+ \epsilon^2 \left(\hat{\mathcal{L}}_{2}^{(L)}[\hat{\rho}(t)]+\hat{\mathcal{L}}_{2}^{(R)}[\hat{\rho}(t)]\right),
\end{align}
where $\hat{\mathcal{L}}_{2}^{(L)}[\hat{\rho}(t)]$ ($\hat{\mathcal{L}}_{2}^{(R)}[\hat{\rho}(t)]$) encodes the effect of the left (right) bath, the full superoperator being $\hat{\mathcal{L}}_{2}[\hat{\rho}(t)]=\hat{\mathcal{L}}_{2}^{(L)}[\hat{\rho}(t)]+\hat{\mathcal{L}}_{2}^{(R)}[\hat{\rho}(t)]$. Using this, we can write down the equation for rate of change of energy in the system,
\begin{align}
	&\frac{d\braket{\hat{H}_S}}{dt}= J_{B_L} +J_{B_R}, \nonumber \\
	& J_{B_L}= \epsilon^2 Tr\left(\hat{H}_S{\mathcal{L}}_{2}^{(L)}[{\rho}(t)]\right),\\ &J_{B_R}= \epsilon^2 Tr\left(\hat{H}_S{\mathcal{L}}_{2}^{(R)}[{\rho}(t)]\right),
	\label{eq:jb}
\end{align}
where $J_{B_L}$ ($J_{B_R}$) can be interpreted as the energy current from the left (right) bath into the system.  

Next, let us look at the relation between the currents from the baths, and the currents in the system.
Due to local conservation of energy, following the continuity equations must be satisfied
\begin{align}
	\label{local_continuity}
	&\frac{d\braket{\hat{H}_L}}{dt}= J_{B_L}-J_{L\rightarrow M}, \nonumber \\
	&\frac{d\braket{\hat{H}_R}}{dt}= J_{B_R}+J_{M\rightarrow R},
\end{align}
where $J_{L\rightarrow M}$ and $J_{M\rightarrow R}$ are the defined in Eq.\eqref{current_defs}. Satisfying this condition requires, along with Eq.\eqref{Local_conservation_condition_on_L}, the following relation is satisfied,
\begin{align}
	\label{superoperator_continuity}
	& Tr\left(\hat{H}_\ell\hat{\mathcal{L}}_{2}^{(L)}[{\rho}(t)]\right)+ Tr\left(\hat{H}_\ell\hat{\mathcal{L}}_{2}^{(R)}[{\rho}(t)]\right)\nonumber \\
	&=Tr\left(\hat{H}_S\hat{\mathcal{L}}_{2}^{(\ell)}[{\rho}(t)]\right),~\ell=L,R. 
\end{align}
From Eqs.(\ref{TCL3}) and (\ref{eq:jb}), we can show that this condition is indeed satisfied and the local continuity equation relating the currents from the bath and the currents inside the system holds.  At NESS, since the rate of change of all the system operators goes to zero, we have
\begin{align}
\label{continuity_of_all_currents}
J_{B_L}=J_{L \rightarrow M} = J_{M \rightarrow R}=-J_{B_R} \geq 0.
\end{align}
This, once again, is a fundamental property of NESS, which is preserved by the RE.

The above discussion is with energy currents, but it holds true for any other currents associated with other local conserved quantities, for example, particle currents.

\subsection{Accuracy of currents from RE}
\label{current_accuracy_RE}
Although the local continuity equations are always satisfied by RE, the currents obtained from RE are only accurate to $O(\epsilon^2)$. This can be seen noting that the currents inside the system are given by coherences in the energy eigenbasis, as shown by Eqs \eqref{current_defs}, \eqref{rate_and_coherence}. To see this explicitly, we write $J_{L\rightarrow M}$ in steady state as
\begin{align}
\label{current_and_coherences}
& J_{L \rightarrow M} = -i\braket{[\hat{H}_M, \hat{H}_L]}=-i\braket{[\hat{H}_S, \hat{H}_L]} \nonumber \\
&= i \sum_{\alpha,\nu=1}^D (E_\alpha - E_\nu) \langle E_\alpha | \hat{\widetilde{{\rho}}}_{\rm NESS} | E_\nu \rangle \langle E_\nu | \hat{H}_L | E_\alpha \rangle \nonumber \\
&=i \epsilon^2\sum_{\alpha,\nu=1}^D (E_\alpha - E_\nu) \langle E_\alpha | \hat{\widetilde{{\rho}}}_{\rm NESS}^{(2)} | E_\nu \rangle \langle E_\nu | \hat{H}_L | E_\alpha \rangle \nonumber\\
&+\textrm{higher orders}.
\end{align}
In the last line, we have used the fact that the leading order coherences are $O(\epsilon^2)$. Since the coherences obtained from RE are given correctly only to $O(\epsilon^2)$, any higher order term obtained from the RE cannot be trusted. Similar expressions can be written for $J_{M \rightarrow R}$ or any other currents related to other local conserved quantities. 

The fact that currents obtained from RE are correct to $O(\epsilon^2)$ can also be seen by calculating the currents from the baths. Carrying out the trace in Eq.\eqref{eq:jb} in the energy eigenbasis of the system, the currents from the baths can be written as
\begin{align}
	\label{current_and_populations}
	J_{B_\ell}&= \epsilon^2 \sum_{\alpha} E_{\alpha}\left\langle E_{\alpha} \left| {\mathcal{L}}_2^{(\ell)}[\hat{\widetilde{{\rho}}}_{\rm NESS}^{(0)}] \right| E_{\alpha} \right\rangle \nonumber \\
	& +\epsilon^4 \sum_{\alpha} E_{\alpha}\left\langle E_{\alpha} \left| \hat{\mathcal{L}}_2^{(\ell)}[\hat{\widetilde{{\rho}}}_{\rm NESS}^{(2)}] \right| E_{\alpha} \right\rangle \nonumber \\
	& + \textrm{higher orders}, \textrm{ where}~\ell=L,R. 
\end{align}
This shows that the leading order currents from the baths are $O(\epsilon^2)$ and depend on $\hat{\widetilde{{\rho}}}_{\rm NESS}^{(0)}$ which, as shown before, is diagonal in the energy eigenbasis and is given correctly by the RE. In other words, the leading order currents from the baths are determined by the leading order populations in the energy eigenbasis. On the other hand, the $O(\epsilon^4)$ term contains contributions from both the diagonal and the off-diagonal elements of  $\hat{\widetilde{{\rho}}}_{\rm NESS}^{(2)}$ in the energy eigenbasis of the system. Since the diagonal elements of $\hat{\widetilde{{\rho}}}_{\rm NESS}^{(2)}$ in the energy eigenbasis of the system are not given correctly by RE, currents from RE will carry $O(\epsilon^4)$ error. 

A particularly critical condition arises when, on physical grounds, the currents are zero, for example, when the temperatures and the chemical potentials of the baths are same, $\beta_L=\beta_R=\beta$, $\mu_L=\mu_R=\mu$. In this case, the $O(\epsilon^2)$ contribution to currents obtained from RE will be zero, since this is given correctly. But, the $O(\epsilon^4)$ term, which can contain an error, may not be zero. This will give unphysical 
$O(\epsilon^4)$ currents even when the current is expected to be zero on physical grounds. Knowing this, it is however possible to clearly identify such spurious unphysical currents stemming from inaccuracies of RE by checking the scaling of the currents with $\epsilon$. If the scaling is $O(\epsilon^2)$, the currents obtained from RE can be trusted. If the scaling in $O(\epsilon^4)$,  it shows that the $O(\epsilon^2)$ contribution to currents is zero. The currents must be taken as zero within the accuracy of RE in that case. This was explicitly used in Ref.\cite{Archak_2020} (see the supplemental material of Ref.\cite{Archak_2020}).

Another interesting point to note from Eqs.\eqref{current_and_coherences},\eqref{current_and_populations} and \eqref{continuity_of_all_currents} is that the $O(\epsilon^2)$ coherences, the $O(\epsilon^0)$ populations in energy eigenbasis and the preservation of local conservation laws are tightly related to one another. If either one of the  populations and the coherences are given accurately, while the other is not, it will lead to violation of local continuity equations.

It is clear from the discussion in this section that the only drawback of the RE is the lack of complete positivity. In the next section, we look in detail at one recent attempt to rigorously rectify this drawback, and show that it violates some of the other required conditions in Table.\ref{table1}. We also make some general comments on all existing weak system-bath coupling Lindblad descriptions.

\section{The Universal Lindblad Equation}
\label{sec:ULE}

As mentioned before, we need to go beyond the standard Lindblad equations in local and eigenbasis forms, LLE and ELE, to satisfy the required conditions in Table.\ref{table1}. Since the RE is microscopically derived and satisfies all other requirements except for the complete positivity, it is intuitive that any QME satisfying the required conditions should give results close to RE. Recently, a Lindblad equation, called the Universal Lindblad Equation (ULE) \cite{ule} has been rigorously derived such that it is of Lindblad form, while maintaining 
\begin{align}
\label{ULE_error}
|| \hat{{\rho}}_{\rm NESS}^{\rm ULE}-\hat{\rho}_{\rm NESS} ||\sim || \hat{\widetilde{{\rho}}}_{\rm NESS}-\hat{\rho}_{\rm NESS} ||\sim O(\epsilon^2),
\end{align}
where $\hat{{\rho}}_{\rm NESS}^{\rm ULE}$ is the NESS density matrix obtained from ULE. The ULE reduces to the LLE and the ELE with further approximations in appropriate limits \cite{ule}. Moreover, as discussed in \cite{ule,Davidovic_2020} it is closely related to other recently derived Lindblad equations to go beyond ELE and LLE. This makes the study of ULE in the light of the fundamental requirements representative of existing Lindblad equation approaches.

\subsection{The general form of ULE}
The ULE approach requires the system-bath coupling to be written in terms of Hermitian operators. So we write the system-bath coupling as, $\hat{H}_{SB}= \sum_\ell^\prime\sum_{k=1,2} \hat{X}_{(\ell, k)} \hat{\mathcal{B}}_{(\ell, k) }$, where
\begin{align}
\label{def_XB}
&\hat{X}_{(\ell, 1)} = \hat{S}_\ell^\dagger + \hat{S}_\ell,~ \hat{X}_{(\ell, 2)} = i(\hat{S}_\ell - \hat{S}_\ell^\dagger),\nonumber \\  
&\hat{\mathcal{B}}_{(\ell, 1)} = \sum_{r=1}^\infty \frac{ \kappa_{r\ell} \hat{B}_{r}^{(\ell) \dagger} + \kappa_{r\ell}^*\hat{B}_{r}^{(\ell)} }{2},\nonumber\\
&\hat{\mathcal{B}}_{(\ell, 2)} = i\sum_{r=1}^\infty  \frac{ \kappa_{r\ell}^* \hat{B}_{r}^{(\ell)} - \kappa_{r\ell} \hat{B}_{r}^{(\ell) \dagger}  }{2}.
\end{align}
The system-bath coupling is then given in the form
\begin{align}
\label{Hermitian_H_SB}
\hat{H}_{SB}= \sum_{\lambda} \hat{X}_{\lambda} \hat{\mathcal{B}}_{\lambda },
\end{align}
where $\lambda=(\ell, k)$ is the combined index.
The ULE is of the form of 
\begin{align}
\label{ule_1}
&\frac{\partial \hat{\rho}}{\partial t}=i[\hat{\rho},\hat{H}_S]+\epsilon^2\hat{\mathcal{L}}_2^\prime[\hat{\rho}]\nonumber \\
&\hat{\mathcal{L}}_2^\prime[\hat{\rho}]=i[\hat{\rho},\hat{H}_{LS}]+\sum_{\lambda}\Big( \hat{L}_{\lambda} \hat{\rho} \hat{L}_{\lambda}^\dagger - \frac{1}{2} \{ \hat{L}_{\lambda}^\dagger \hat{L}_{\lambda}, \hat{\rho}   \} \Big),
\end{align}
with  the Lindblad operators and the Lamb-shift Hamiltonian being given by 
\begin{equation}
\label{ule_2}
\begin{aligned}
&\hat{L}_{\lambda} =  \sum_{\lambda^\prime} \int_{-\infty}^{\infty} \hspace{-1em} ds \widetilde{\bm{g}}_{\lambda, \lambda^\prime} (s) \hat{X}_{\lambda^\prime}(-s) \\
&\hat{H}_{\rm LS}=\frac{1}{2i} \int_{-\infty}^{\infty} \hspace{-1em} ds \int_{-\infty}^{\infty} \hspace{-1em} ds^\prime \sum_{\lambda,\lambda^\prime} \hat{X}_{\lambda} (s) \hat{X}_{\lambda^\prime}(s^\prime) \bm{\phi}_{\lambda \lambda^\prime}(s,s^\prime)  
\end{aligned}
\end{equation}
where $\hat{X}_\lambda(t)=e^{i \hat{H}_S t} \hat{X}_\lambda e^{-i \hat{H}_S t}$ denotes the interaction picture operator. The matrices $\bm{\phi}(t,s)$ and $\widetilde{\bm{g}}(t)$ are,
\begin{align}
\widetilde{\bm{g}}(t)&=\int_{-\infty}^{\infty} \hspace{-0.8em} d\omega \bm{g}(\omega) e^{-i \omega t},~\bm{\phi}(t,s)=\widetilde{\bm{g}}(t) \widetilde{\bm{g}}(-s) \text{sgn} (t-s), \nonumber \\
\bm{g}(\omega)&=\sqrt{\frac{\bm{G}(\omega)}{2\pi}},~\bm{G}_{\lambda,\lambda^\prime}(\omega)= \int_{-\infty}^{\infty} \frac{dt}{2 \pi} \braket{\hat{\mathcal{B}}_\lambda (t) \hat{\mathcal{B}}_{\lambda^\prime}(0)}_B  e^{i \omega t}
\end{align}
with $\text{sgn}(x)$ being the sign function, and $\braket{...}_B$ denoting the expectation value taken over the bath initial state. Given this general form the ULE we now investigate if the ULE satisfies the fundamental requirements in Table~\ref{table1}.

\subsection{Accuracy of ULE}

\subsubsection{Accurate populations but inaccurate coherences}

As mentioned before, the ULE has been derived so that the steady state it predicts differs from the exact steady state in $O(\epsilon^2)$. This means that the $O(\epsilon^0)$ term of the NESS density matrix, must be given correctly by the ULE, and hence would match with that from RE. Here we explain how to see this explicitly in general.

For this, we  write $\hat{\widetilde{\rho}}_{\rm NESS}^{(0)}$ in the energy eigenbasis, as in Eq.(\ref{rho_NESS_0}) and explicitly evaluate the term required by the first condition in Table~\ref{table1},
\begin{align}
\label{explicit_ULE_LE_condition}
&\left\langle E_{\alpha} \left| \hat{\mathcal{L}}_2^\prime[\hat{\widetilde{{\rho}}}_{\rm NESS}^{(0)}] \right| E_{\alpha} \right\rangle\nonumber =\\
&\sum_{\nu, \lambda , \lambda^\prime} 2\pi p_{\nu}\Big\{  \bra{E_\alpha} \hat{X}_{\lambda^\prime} \ket{E_\nu}\bra {E_\nu} \hat{X}_{\lambda} \ket{E_\alpha}  \bm{G}_{\lambda,\lambda^\prime}(E_\nu - E_\alpha)  \nonumber \\
&-\sum_{\gamma} \bra {E_\alpha} \hat{X}_{\lambda} \ket{E_\gamma}\bra{E_\gamma} \hat{X}_{\lambda^\prime} \ket{E_\alpha}  \bm{G}_{\lambda,\lambda^\prime}(E_\alpha - E_\gamma)\delta_{\alpha \nu} \Big\}.
\end{align}
We need to compare this result with that obtained from RE to check the first condition in Table~\ref{table1}. To do this, it is easier to write the RE also in the form where the system-bath coupling terms are Hermitian operators, as in Eq.(\ref{Hermitian_H_SB}). The RE for this form of coupling is given by
\begin{align}
\label{Hermitian_RE}
&\frac{\partial\hat{\rho}}{\partial t} = i[\hat{\rho},\hat{H}_S]- \epsilon^2 \sum_{\lambda,\lambda^\prime} \Big ( \big[\hat{X}_\lambda, \hat{X}_{\lambda^\prime \lambda} \hat{\rho}\big] + {\rm h.c.} \Big),
\end{align}
with
\begin{align}
& \hat{X}_{\lambda^\prime \lambda}=\int_0^\infty dt^\prime \langle \hat{\mathcal{B}}_\lambda(t^\prime) \hat{\mathcal{B}}_{\lambda^\prime}(0) \rangle \hat{X}_{\lambda^\prime}(-t^\prime), \nonumber \\
&=\int_0^\infty dt^\prime \int_{-\infty}^{\infty}\bm{G}_{\lambda,\lambda^\prime}(\omega)e^{-i\omega t^\prime} \hat{X}_{\lambda^\prime}(-t^\prime).
\end{align}
Using Eq.(\ref{def_XB}), it can be checked, that Eq.(\ref{Hermitian_RE}) can be reduced to Eq.(\ref{RE2}).
With Eq.(\ref{Hermitian_RE}), direct evaluation of $\left\langle E_{\alpha} \left| \hat{\mathcal{L}}_2[\hat{\widetilde{{\rho}}}_{\rm NESS}^{(0)}] \right| E_{\alpha} \right\rangle$ gives exactly the same expression as Eq.(\ref{explicit_ULE_LE_condition}). Thus the ULE explicitly satisfies the condition first condition in Table~\ref{table1}. It is important to note that we do not explicitly require the values of the populations $p_\nu$ to show that the condition is satisfied.

Using exactly similar steps, it can be checked that the ULE does not satisfy the second condition in Table~\ref{table1}. Thus, the leading order coherences from ULE are not given correctly. Since the leading order coherences in energy eigenbasis are $O(\epsilon^2)$, this is consistent with the ULE having an error of $O(\epsilon^2)$ (see Eq.\eqref{ULE_error}).

\begin{figure*}
\includegraphics[width=\textwidth]{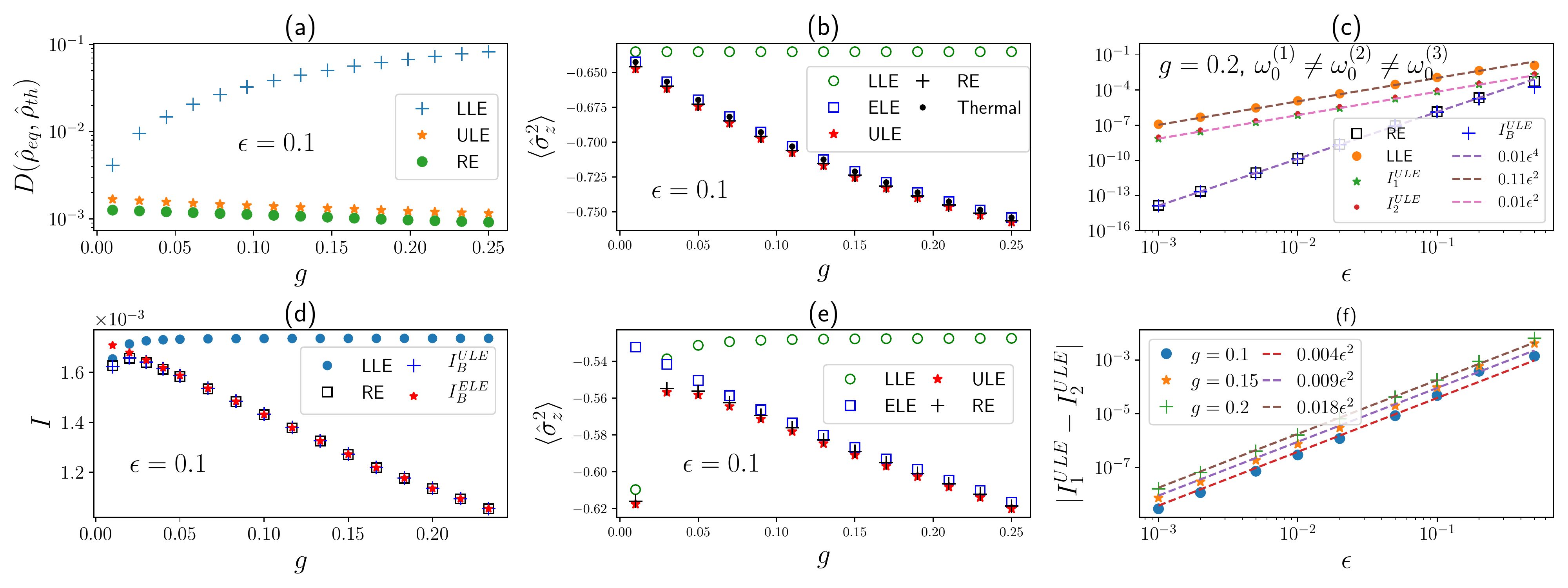}
\caption{\label{fig:mainall_plots} The top row [(a), (b), (c)] is for the equilibrium case ($\beta_L=1$, $\beta_R=1$) and the bottom row [(d), (e), (f)] is for the non-equilibrium case ($\beta_L=5$, $\beta_R=0.5$). Apart from (c), in all other cases, $\omega_0^{(1)}=\omega_0^{(2)}=\omega_0^{(3)}=1$. {\bf (a)} The trace distance between the expected thermal state $\hat{\rho}_{\rm th}$ and the equilibrium  steady-state $\hat{\rho}_{\rm eq}$  as obtained by the LLE, ULE and RE as a function of $g$. {\bf (b)} The local magnetization in equilibrium steady state as obtained by LLE, ULE, RE, ELE and by the expected thermal state as a function of $g$. {\bf (c)} The scaling of spin currents in equilibrium as given by LLE, ULE, RE as a function of system-bath coupling strength $\epsilon$, for the case where $\omega_0^{(1)}=1, \omega_0^{(2)}=1.5, \omega_0^{(3)}=2$. Here $I_1^{ULE}$, $I_2^{ULE}$ and $I_B^{ULE}$ refer to the two bond-currents and the boundary current as obtained by ULE. {\bf (d)} Spin currents in NESS  as obtained by LLE, ULE, RE and ELE. For ULE and ELE, only the boundary currents $I_B^{ULE}$ and $I_B^{ELE}$ are plotted. (For ELE, the bond currents are exactly zero.) {\bf (e)} The local magnetization in NESS as obtained by LLE, ULE, RE, ELE as a function of $g$. {\bf (f)} The scaling of difference between the two bond currents at NESS from ULE as a function of system-bath coupling strength $\epsilon$.  Other parameters: $\mu_L=\mu_R=-0.5$, $\omega_c=10$. All energy parameters are in units of $\omega_0^{(1)}$.   }
\end{figure*}

\subsubsection{Violation of local conservation laws}

As we have already discussed, having correct populations to leading order while having incorrect coherences leads to violation of local conservation laws. So, the ULE does not satisfy the fourth requirement in Table.~\ref{table1}. 
This can be explicitly seen by writing the evolution equation for any system operator, 
\begin{align}
\label{operator_ULE}
\frac{d\braket{\hat{O}}}{dt}&=-i\braket{[\hat{O},\hat{H}_S+\hat{H}_{\rm LS}]}\nonumber \\
& +\sum_\lambda \frac{1}{2}\left( \braket{[\hat{L}_\lambda^\dagger,\hat{O}] \hat{L}_\lambda} + \braket{\hat{L}_\lambda^\dagger [\hat{O} ,\hat{L}_\lambda] } \right). 
\end{align}
From the form of the operators $\hat{L}_\lambda$ in Eq.\eqref{ule_2}, we see that even if $[\hat{O},\hat{H}_{SB}]=0$, ${\rm Tr}\left( \hat{O} \hat{\mathcal{L}}_2^\prime(\hat{\rho}) \right) \neq 0$.

\subsubsection{Accuracy of currents from ULE}
\label{current_accuracy_ULE}
Despite this violation of local conservation laws, the ULE gives the correct currents from the baths to the leading order. This can be seen by writing the ULE in the form of Eq.\eqref{TCL3} and defining the corresponding currents as in Eq.\eqref{eq:jb}. The currents can then be written in energy eigenbasis, as in Eq.\eqref{current_and_populations}, to see that the leading order contributions are given by the $O(\epsilon^0)$ populations in the energy eigenbasis. Since these terms are given accurately by the ULE,  it follows that the currents from the baths are given accurately to $O(\epsilon^2)$ and agree with the same currents obtained from RE. The error in the currents from the baths as obtained from ULE is therefore $O(\epsilon^4)$, which is same as that from RE. 

However, since the coherences in energy eigenbasis contain $O(\epsilon^2)$ error in ULE, the currents inside the system, defined in Eq.\eqref{current_defs}, will have an error in the leading order. This, once again, shows that local conservation laws will be violated.

\subsection{General comments regarding Lindblad equations}
We end this section with some general comments regarding Lindblad equations. We have already mentioned that the ULE is closely related to some of the other Lindblad equations devised to go beyond the limitations of ELE and LLE. It is worth mentioning that, to our knowledge, all existing Lindblad equations \cite{ule,Kleinherbers_2020,Davidovic_2020,mozgunov2020,mccauley2020,
kirvsanskas2018} except the LLE violate the  fourth condition in Table.~\ref{table1}, thereby violating local conservation laws. The LLE on the other hand is known to not satisfy the thermalization requirement, the third condition in Table~\ref{table1}. Further, although explicitly satisfying the local conservation laws, it does not always give the correct currents in NESS \cite{Walls1970,Wichterich_2007,Rivas_2010,barranco_2014,Levy2014,archak,
Trushechkin_2016,
Eastham_2016,Hofer_2017,Gonzalez_2017,Mitchison_2018,
Cattaneo_2019,Hartmann_2020_1,Benatti_2020,konopik_2020local,
Scali_2021,Floreanini_2021,trushechkin2021}. This is contrary to the ULE, which despite violating local conservation laws, gives correct currents from the baths to the leading order. The RE, on the other hand, except for generically violating the requirement of complete positivity, satisfies all other the conditions in Table~\ref{table1}. It therefore seems that a weak system-bath coupling QME respecting all the conditions in Table~\ref{table1} is generically impossible. In particular, since currents inside the system are related to coherences and currents from the baths are related to populations, it seems that, no weak system-bath coupling QME can generically satisfy the requirement of complete positivity and simultaneously give correct populations and coherences to the leading order. Thus all weak system-bath coupling Lindblad equations seem fundamentally limited. While this is true for the existing QMEs to our knowledge, a general proof of the fact is missing till now. We leave the general proof of this to future work. Although all the results above have been derived keeping a set-up of the form of Fig.~\ref{fig:maintwo_terminal_schematic} in mind, these can be easily carried forward to cases where there are more than two baths.

Our discussion till now has been general. We have not yet written down any particular system Hamiltonian, or chosen any particular bath spectral function. This completely general discussion of accuracy of QMEs sets apart our work from most previous works\cite{Walls1970,Wichterich_2007,Rivas_2010,barranco_2014,Levy2014,archak,
Trushechkin_2016,
Eastham_2016,Hofer_2017,Gonzalez_2017,Mitchison_2018,
Cattaneo_2019,Hartmann_2020_1,Benatti_2020,konopik_2020local,
Scali_2021,Floreanini_2021,trushechkin2021}, where the accuracy is discussed  referring to a particular model. In the following, we numerically check our general discussion in the three site Heisenberg model coupled to bosonic baths.

\begin{figure*}[t]
	\includegraphics[width=\textwidth]{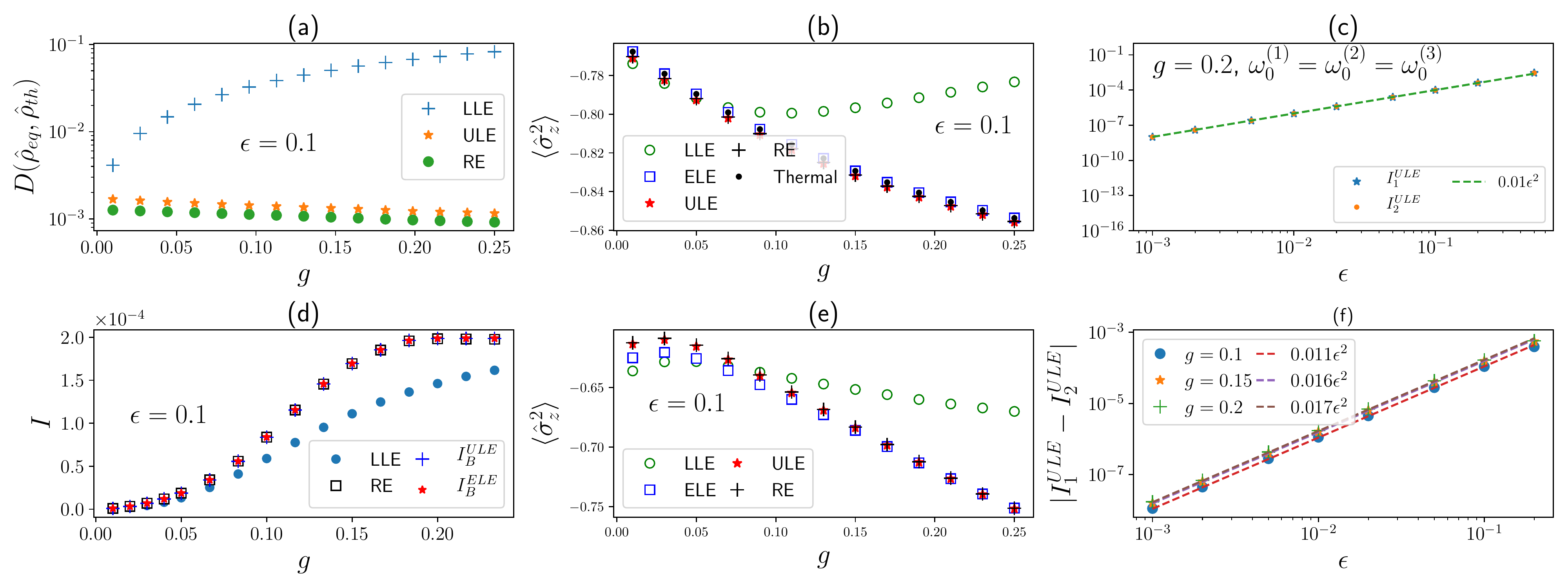}
	\caption{\label{fig:all_plots2} The top row [(a), (b), (c)] is for the equilibrium case ($\beta_L=1$, $\beta_R=1$) and the bottom row [(d), (e), (f)] is for the non-equilibrium case ($\beta_L=5$, $\beta_R=0.5$). Apart from (c), everywhere, $\omega_0^{(1)}=1, \omega_0^{(2)}=1.5, \omega_0^{(3)}=2$. {\bf (a)} The trace distance between the expected thermal state $\hat{\rho}_{th}$ and the equilibrium steady-state $\hat{\rho}_{eq}$  as obtained by the local Lindblad (LLE), Universal Lindblad (ULE) and and Redfield (RE) equations as a function of $g$. {\bf (b)} The local magnetization in equilibrium steady state as obtained by LLE, ULE, RE, ELE and by the expected thermal state as a function of $g$. {\bf (c)} The scaling of bond spin currents in equilibrium as obtained from ULE, $I_1^{ULE}$, $I_2^{ULE}$, for the case where $\omega_0^{(1)}=\omega_0^{(2)}=\omega_0^{(3)}=1$. The LLE, RE and ULE boundary currents are zero correct to numerical precision of $10^{-16}$. {\bf (d)} Spin currents in NESS as obtained by LLE, ULE, RE and ELE. For ULE and ELE, the boundary currents $I_B^{ULE}$ and $I_B^{ELE}$ are plotted. {\bf (e)} The local magnetization in NESS as obtained by LLE, ULE, RE, ELE as a function of $g$. {\bf (f)} The scaling of difference between the two bond currents at NESS from ULE as a function of system-bath coupling strength $\epsilon$.   Other parameters: $\mu_1=\mu_2=-0.5$, $\omega_c=10$. All energy parameters are in units of $\omega_0^{(1)}$.   }
\end{figure*}

\section{Numerical Results}
\label{sec:numerical}

We now numerically exemplify the above discussion by using a XXZ spin-chain in the presence of a magnetic field, with the first and the last sites attached to baths modelled by infinite number of bosonic modes, 
\begin{align}
& \hat{H}_S = \sum_{\ell=1}^N \frac{\omega^{(\ell)}_0}{2} \hat{\sigma}_z^\ell  -  \sum_ {\ell=1}^{N-1} g(\hat{\sigma}_x^\ell \hat{\sigma}_x^{\ell+1} + \hat{\sigma}_y^\ell \hat{\sigma}_y^{\ell+1} + \Delta \hat{\sigma}_z^\ell \hat{\sigma}_z^{\ell+1}), \nonumber\\
& \hat{H}_{SB}  = \sum_{\ell=1,N}\sum_{r=1}^\infty  (\kappa_{\ell r} \hat{B}^{ {(\ell)} \dagger}_r \hat{\sigma}^\ell_- + \kappa_{lr}^* \hat{B}^{(\ell)}_r \hat{\sigma}^\ell_+),
\end{align} 
where $\hat{\sigma}^\ell_{x,y,z}$ denotes the Pauli matrices acting on the $\ell^{\text{th}}$ spin, $\hat{\sigma}^\ell_{+}=(\hat{\sigma}^\ell_{x}+i \hat{\sigma}^\ell_{y})/2$, $\hat{\sigma}^\ell_{-}=(\hat{\sigma}^\ell_{x}-i \hat{\sigma}^\ell_{y})/2$, $\hat{B}^{(\ell)}_r$ is bosonic annihilation operator for the $r^{\text{th}}$ mode of the bath attached at the $\ell^{\text{th}}$ site. Here, $\omega^{(\ell)}_0$, $g$, and $g\Delta$ represent the magnetic field, the overall spin-spin coupling strength and the anisotropy respectively. The total number of excitations in the system is given by $\hat{N}_{S}=\sum_{\ell=1}^N \hat{\sigma}^\ell_{+} \hat{\sigma}^\ell_{-}$ and it satisfies $[\hat{N}_{S},\hat{H}_S]=0 $.  We consider bosonic baths described by Ohmic spectral functions with Gaussian cut-offs, $ \mathfrak{J}_\ell(\omega)=\sum_k 2 \pi \left| \kappa_{\ell k} \right|^2 \delta (\omega-\Omega^\ell_k)=\omega e^{-(\omega / \omega_c)^2}\Theta(\omega),$ where, $\Theta(\omega)$ is Heaviside step function, and $\omega_c$ is the cut-off frequency. The set-up is exactly of the form of Fig.~\ref{fig:maintwo_terminal_schematic}, with the initial inverse temperatures and chemical potentials of the baths as shown. We look at the equilibrium and non-equilibrium steady states as obtained by RE, LLE, ELE, and ULE. The explicit forms of all these equations for the XXZ spin-chain are given in the Appendix (Appedices.~\ref{appendix:RE},~\ref{appendix:LLE},~\ref{appendix:ELE} and ~\ref{appendix:ULE} respectively). For numerical simplicity, we consider the $N=3$ case, which is sufficient to demonstrate the fundamental limitations of all the three Lindblad equations.    All numerical results below are obtained using QuTiP \cite{qutip_1,qutip_2}.

\begin{table*}
	\setlength{\tabcolsep}{14pt}
	\begin{tabular}{|c|c|c|c|c|} 
		\hline
		& LLE & ELE & ULE & RE \\[0.5ex] 
		\hline
		Diagonal (Populations) & O(1) wrong & O(1) wrong in NESS  & O(1) correct & O(1) correct \\ 
		\hline
		Off-diagonal (Coherences) & O($\epsilon^2$) wrong & O($\epsilon^2$) wrong & O($\epsilon^2$) wrong & O($\epsilon^2$) correct \\
		\hline
		Conservation Laws & Conserved & Violated & Violated & Conserved \\
		\hline
		Thermalization & Violated & Preserved & Preserved & Preserved  \\
		\hline
		Currents from baths & valid for $g\ll \epsilon$ & valid for $|E_{\alpha}-E_{\gamma}| \gg \epsilon$   & valid for all $g$  &  valid for all $g$ \\ 
		\hline 
		Currents in system & valid for $g\ll \epsilon$ & zero   & O($\epsilon^2$) wrong &   O($\epsilon^2$) correct \\
		\hline
		Complete positivity & Preserved & Preserved & Preserved & Violated \\
		\hline
	\end{tabular}
	\label{table}
	\caption{Accuracy of various QMEs in various settings.\label{Table:accuracies}}
\end{table*}

First, we look at the equilibrium case (top row of Fig.~\ref{fig:mainall_plots}), $\beta_L=\beta_R=\beta$, $\mu_L=\mu_R=\mu$, and calculate the trace distance \cite{NnC}, $D(\hat{\rho}_{\rm eq},\hat{\rho}_{\rm th})={\rm Tr}[\sqrt{(\hat{\rho}_{\rm eq}-\hat{\rho}_{\rm th})^2}]/2$, between the steady state $\hat{\rho}_{\rm eq}$ given by the three QMEs and the thermal state $\hat{\rho}_{\rm th}~=~\frac{e^{-\beta(\hat{H}_S-\mu \hat{N}_S)}}{{\rm Tr}\left( e^{-\beta(\hat{H}_S-\mu \hat{N}_S)}\right)}$. The trace distance $D(\hat{\rho}_{\rm eq},\hat{\rho}_{\rm th})$ is plotted as a function of $g$ for $\epsilon=0.1$ in Fig.~\ref{fig:mainall_plots}(a). We see that the trace distance is quite small and of the same order for RE and ULE, and does not vary much with $g$. On the other hand, the trace distance for LLE is larger and grows with $g$ while that for ELE is identically zero.  
Note that generically any system is expected to have steady-state coherences in energy eigenbasis in equilibrium \cite{Guarnieri_2018,Archak_2020} for any finite $\epsilon$ and convergence to $\hat{\rho}_{\rm th}$ is physically only expected for $\epsilon\rightarrow 0$. Nevertheless, as shown in Fig~\ref{fig:mainall_plots}(b), observables like local magnetizations, $\braket{\hat{\sigma}^\ell_z}$, obtained from ULE and RE show very small difference from the thermal expectation values, since they depend on populations in leading order.  

The same is not true for the spin currents. The bond spin currents in the system $I_j$ are defined from the continuity equation
$
\frac{d\braket{\hat{\sigma}^j_z}}{dt} =-i\braket{[\hat{\sigma}^j_z,\hat{H}_S]}= I_{j} - I_{j-1}
$, which gives 
$
I_j=4i g(\braket{\hat{\sigma}^j_+ \hat{\sigma}^{j+1}_-} - \braket{\hat{\sigma}^j_- \hat{\sigma}^{j+1}_+}).
$ 
On the other hand, boundary spin-currents are defined by the continuity equation,  $\frac{d\braket{\hat{M}_z}}{dt}= I_{B_L} +I_{B_R}$, where $\hat{M}_z=\sum_{\ell=1}^N \hat{\sigma}^\ell_z$. For equations of the form Eq.(\ref{TCL3}), $I_{B_\ell}= \epsilon^2 Tr\left(\hat{M}_z\hat{\mathcal{L}}_{2}^{(\ell)}[\hat{\rho}(t)]\right)$, $\ell=L,R$ where $\hat{\mathcal{L}}_{2}^{(\ell)}$ changes depending on the QME used. In steady state, local conservation laws require $I_1= - I_{B_L}=I_2= I_{B_R}$. Further, in equilibrium, currents should be zero \cite{Levy2014}. 
We find that, for $\omega_0^{(1)}\neq\omega_0^{(2)}\neq \omega_0^{(3)}$, all QMEs except ELE gives unphysical currents in even though we have $\beta_L=\beta_R=\beta$, $\mu_L=\mu_R=\mu$. However, it is paramount to note that, as shown in  Fig~\ref{fig:mainall_plots}(c),  the currents from RE and the boundary currents from ULE scale as $\epsilon^4$, clearly showing that the leading order, $O(\epsilon^2)$, term is zero. This is completely consistent with our discussion in Sec.~\ref{current_accuracy_RE} and Sec.~\ref{current_accuracy_ULE}. On the other hand, the bond currents from ULE are different, and scale as $\epsilon^2$, thereby showing error in leading order, and violating local conservation laws. The LLE does not violate local conservation laws, but still gives equilibrium currents scaling as $\epsilon^2$ \cite{Levy2014}, thereby showing leading order inaccuracies in both populations and coherences. 

Now we discuss the results for the non-equilibrium set-up (bottom row of Fig.~\ref{fig:mainall_plots}), $\beta_L \neq \beta_R$. 
The currents in NESS as a function of $g$ are shown in Fig.~\ref{fig:mainall_plots}(d). The current from RE matches the boundary current from ULE for all $g$ and shows a non-monotonic behavior with $g$. On the other hand, LLE fail to capture the non-monotonicity of current as a function of $g$, and matches only at very small values of $g$. The ELE identically gives zero  currents inside the system, even in NESS \cite{Wichterich_2007}. However, the boundary currents from ELE seem to match reasonably well with RE for sufficiently high $g$. Thus, the ELE also violates the local conservation laws in NESS. The local magnetizations in NESS also match from RE and ULE and are significantly different from the LLE and ELE results at high and low $g$ respectively, as shown in Fig.~\ref{fig:mainall_plots}(e). Finally, in Fig.~\ref{fig:mainall_plots}(f), we demonstrate that ULE bond currents are different in NESS with the difference scaling as $O(\epsilon^2)$, which highlights a clear violation of local conservation laws in ULE.

All plots in Fig.~\ref{fig:mainall_plots} are made with $\omega_0^{(1)}=\omega_0^{(2)}=\omega_0^{(3)}=1$, except for panel (c)  which was plotted for $\omega_0^{(1)}=1, \omega_0^{(2)}=1.5, \omega_0^{(3)}=2$. In Fig.~\ref{fig:all_plots2}, we give the complimentary plot, where all numerical results execpt panel (c) are for $\omega_0^{(1)}=\omega_0^{(2)}=\omega_0^{(3)}=1$, while panel (c) is for $\omega_0^{(1)}=1, \omega_0^{(2)}=1.5, \omega_0^{(3)}=2$.  Two important things are worth noticing. First, when $\omega_0^{(1)}=\omega_0^{(2)}=\omega_0^{(3)}$, none of LLE,  RE and ULE boundary currents give any unphysical non-zero result (correct to a numerical precision of $10^{-16}$) if $\beta_L=\beta_R=\beta$, $\mu_L=\mu_R=\mu$. However, the ULE bond currents still gives non-zero $O(\epsilon^2)$ result and thereby violate local conservation laws. Second, for  $\omega_0^{(1)}=1, \omega_0^{(2)}=1.5, \omega_0^{(3)}=2$, because of already large separation between the on-site magnetic fields, the secular approximation performs much better, and as a result, ELE gives reasonably accurate results both in equilibrium and out-of-equilibrium steady states, for a wide range of $g$.  However, ELE, by construction, gives zero currents inside the system \cite{Wichterich_2007}.  Table \ref{Table:accuracies} summarizes the accuracies and validity regimes of the various QMEs.  Although our numerical demonstration is for $N=3$, our analytical understanding in previous sections shows that these issues persist for larger $N$, as well as generically for other systems.

\section{Summary and discussions}
\label{sec:summary_and_discussions}

In this work, we have investigated fundamental limitations of QMEs obtained in weak system-bath coupling approach to describe a quantum system coupled to multiple macroscopic baths which can be at different temperatures and chemical potentials, in absence of any external driving (i.e, the when the Hamiltonian is time-independent). Microscopically obtaining the QME up to leading order in system-bath coupling leads to the so-called Redfield form, which is known to generically violate the requirement of complete positivity. Very often further approximations are done on the Redfield form to reduce it to a Lindblad form, so that complete positivity is restored. We have laid down some fundamental requirements (Table.~\ref{table1}) that any such approximation must satisfy. If they are violated there occurs physical inconsistencies like inaccuracies in leading order populations and coherences in the energy eigenbasis, violation of thermalization when all baths have same temperatures and chemical potentials, violation of local conservation laws at NESS. This has allowed us to check for occurrence of these physical inconsistencies in weak-system-bath coupling QMEs in general, without writing them down for any particular model. This model independent general discussion distinguishes our work from most previous works investigating accuracies of various QMEs \cite{Walls1970,Wichterich_2007,Rivas_2010,barranco_2014,Levy2014,archak,
Trushechkin_2016,
Eastham_2016,Hofer_2017,Gonzalez_2017,Mitchison_2018,
Cattaneo_2019,Hartmann_2020_1,Benatti_2020,konopik_2020local,
Scali_2021,Floreanini_2021}.

We have found that the Redfield equation, despite being generically not completely positive, does not violate any other of the fundamental requirements. It therefore gives correct populations and coherences in energy eigenbasis to the leading order, shows thermalization, and preserves local conservation laws. On the other hand, weak system-bath coupling Lindblad descriptions violate one or more the fundamental requirements and thereby have one or more of the physical inconsistencies mentioned above, despite being completely positive. In particular, it can be argued that no existing weak system-bath coupling Lindblad description, to our knowledge, can generically give both correct populations and correct coherences to the leading order. As an example,  we have explicitly shown these violations in generality for the so-called universal Lindblad equation which has been recently derived. We have also numerically demonstrated these statements in a three site XXZ chain coupled to two bosonic baths. From our results, it seems that there is no consistent way to correct the violation of complete positivity of the Redfield equation, without going to higher orders in system-bath coupling. 

These results are extremely significant because weak system-bath coupling Lindblad equations remain the most widely-used descriptions of open quantum systems, and accurate description of populations, coherences and currents is important for quantum information and thermodynamics  \cite{NnC,Bennett2000,Streltsov_2017,Lostaglio_2015,Narasimhachar_2015,
Mitchison_2015,Allahverdyan_2004,Korzekwa_2016,Kammerlander_2016,
Santos_2019,Francica_2019}. It seems the Redfield equation, despite not being completely positive, should be the description of choice. The various errors in the Redfield equations are controlled. Whether a result obtained from the Redfield equation can be trusted or not, can be determined by checking its scaling with system-bath coupling strength. Further, there have been techniques suggested to infer the next order corrections to populations in steady-state from Redfield equation without actually writing down the next higher order QME \cite{Juzar_2012, Juzar_2013, Juzar_2017}. These techniques may be able to correct the positivity issues in the steady-state obtained from the Redfield equation \cite{Archak_2020}.

It is paramount to mention that the fundamental limitations of Lindblad descriptions discussed in this paper pertain to the case of a system weakly coupled to multiple thermal baths in absence of any external driving. In presence of particular kinds of driving, Lindblad equations can be microscopically derived even for strong system-bath coupling \cite{collision1}. Our results do not pertain to those descriptions. Further, our results do not pertain to cases where the Lindblad dissipators do not act on the system, but rather on the bath degrees of freedom. Such approaches can be shown to accurately describe the steady state of the system for arbitrary strength of system-bath coupling \cite{thermleads}. Our results, however, do show that it is imperative to go beyond weak system-bath coupling QMEs for physically consistent description of open quantum systems.

\section*{Acknowledgements}
\label{appendix_1}
 MK would like to acknowledge support from the project 6004-1 of the Indo-French Centre for the Promotion of Advanced Research (IFCPAR), Ramanujan Fellowship (SB/S2/RJN-114/2016), SERB Early Career Research Award (ECR/2018/002085) and SERB Matrics Grant (MTR/2019/001101) from the Science and Engineering Research Board (SERB), Department of Science and Technology, Government of India. AD and MK acknowledge support of the Department of Atomic Energy, Government of India, under Project No. RTI4001. AP acknowledges funding from the European Research Council (ERC) under the European Unions Horizon 2020 research and innovation program (Grant Agreement No. 758403).

\appendix
\section*{Appendix}

The example we looked at in the main text is the three-site Heisenberg model, coupled to two different bosonic baths at the first and the third sites. Here, we will give the QMEs used for this set-up, and write them for Heisenberg spin-chain of length $N$.

\section{Redfield equation}
\label{appendix:RE}

The Redfield equation for our set-up is given by  
\begin{equation}
	\begin{aligned}
		\frac{\partial \hat{\rho}}{\partial t}  & =i [\hat{\rho}, \hat{H}_S]  \\
		- \epsilon ^2   &\sum_{\ell=1,N}\sum_{\alpha, \gamma=1}^{2^N}  \Big\{   \left[ \hat{\rho}  \ket{E_\alpha} \bra{E_\alpha} \hat{\sigma}^\ell_-  \ket{E_\gamma} \bra{E_\gamma} ,  \hat{\sigma}^l_+   \right]   C_{\ell}(\alpha,\gamma)\\ 
		+&\left[\hat{\sigma}^\ell_+  ,  \ket{E_\alpha} \bra{E_\alpha} \hat{\sigma}^l_-  \ket{E_\gamma}\bra{E_\gamma} \hat{\rho} \right]  D_{\ell}(\alpha,\gamma) 
		+\text{H.c}  \Big\} 
		\label{redfield_final}
	\end{aligned}
\end{equation}
with
\begin{equation}
	\begin{aligned}
		C_{\ell}(\alpha,\gamma) &= \frac{\mathfrak{J}_{\ell}(E_{\gamma \alpha}) n_{\ell}(E_{\gamma \alpha})}{2 } - i \mathcal{P} \int_{0}^{\infty} d \omega \frac{\mathfrak{J}_{\ell}(\omega) n_{\ell}(\omega)}{\omega-E_{\gamma \alpha}},  \\
		D_{\ell}(\alpha,\gamma) &= \frac{ e^{\beta_{\ell}(E_{\gamma \alpha}-\mu_{\ell})} \mathfrak{J}_{\ell}(E_{\gamma \alpha}) n_{\ell}(E_{\gamma \alpha})}{2 }\\ &- i \mathcal{P} \int_{0}^{\infty} d \omega \frac{e^{\beta_{\ell}(\omega-\mu_{\ell})} \mathfrak{J}_{\ell}(\omega) n_{\ell}(\omega)}{\omega-E_{\gamma \alpha}},
		\label{redfield:constants}
	\end{aligned}
\end{equation}
where $\mathcal{P}$ denotes the Cauchy Principal value. 
Here $\ket{E_\alpha}$ and $\ket{E_\gamma}$ are simultaneous eigenkets of the system hamiltonian and the magnetization operator, with eigenenergies $E_\alpha$ and $E_\gamma$, where $E_{\gamma \alpha} = E_\gamma-E_\alpha$.

\section{Local Lindblad equation}
\label{appendix:LLE}
The local-Lindblad equation, which can be derived microscopically for $g\ll\epsilon$, from the Redfield equation, is given by,
\begin{equation}
	\begin{aligned}
		\frac{\partial \hat{\rho}}{\partial t} =& i  [ \hat{\rho},\hat{H}+\hat{H}_{LS} ]  \\
		+ \epsilon^2 & \sum_{\ell=1,N} \mathfrak{J}_\ell(\omega^{(\ell)}_0)(n_\ell(\omega^{(\ell)}_0) + 1) \left( \hat{\sigma}^\ell_- \hat{\rho} \hat{\sigma}^\ell_+ - \frac{1}{2} \{ \hat{\sigma}^\ell_+ \hat{\sigma}^\ell_-,\hat{\rho} \}  \right) \\ + &  \mathfrak{J}_\ell(\omega^{(\ell)}_0) n_\ell(\omega^{(\ell)}_0) \left(  \hat{\sigma}^\ell_+ \hat{\rho} \hat{\sigma}^\ell_- - \frac{1}{2} \{ \hat{\sigma}^\ell_- \hat{\sigma}^\ell_+, \hat{\rho} \} \right) \label{locallindblad}
	\end{aligned}
\end{equation}
where $\hat{H}_{LS}= \sum_{\ell=1,N}(\overline{\Delta}_\ell + 2\overline{\Delta}^\prime_\ell)\frac{\omega^{(\ell)}_0}{2}$, and   $\overline{\Delta}_\ell$ and $\overline{\Delta}_\ell^\prime$ are given by, 
\begin{equation}
	\begin{aligned}
		\overline{\Delta}_\ell &=\mathcal{P} \frac{\epsilon^2}{2\pi}\int_{0}^{\infty}d\omega \frac{ \mathfrak{J}_\ell(\omega)}{\omega^{(\ell)}_0 -\omega}, \\
		\overline{\Delta}_i^\prime &= \mathcal{P} \frac{\epsilon^2}{2\pi} \int_{0}^{\infty}d\omega \frac{ \mathfrak{J}_\ell(\omega) n_\ell(\omega)}{\omega^{(\ell)}_0 -\omega}.
	\end{aligned}
\end{equation}

\section{Eigenbasis Lindblad equation}
\label{appendix:ELE}
The eigenbasis Lindblad equation is obtained from the Redfield equation by the doing the so-called secular approximation \cite{breuer_book}. This amounts to an equation of the form
\begin{equation}
	\begin{aligned}
		\frac{\partial \hat{\rho}}{\partial t}  & =i [\hat{\rho}, \hat{H}_S] - \epsilon ^2   \sum_{\ell=1,N}\sum_{\alpha, \gamma=1}^{2^N}  \sum_{ \substack{\alpha^\prime,\gamma^\prime=1 \\ E_{\gamma^\prime \alpha^\prime}=E_{\gamma \alpha}}}^{2^N}\\  \Big\{  \big[ & \hat{\rho}  \ket{E_\alpha} \bra{E_\alpha} \hat{\sigma}^\ell_-  \ket{E_\gamma} \bra{E_\gamma} , \ket{E_{\gamma^\prime}} \bra{E_{\gamma^\prime}} \hat{\sigma}^\ell_+    \ket{E_{\alpha^\prime}} \bra{E_{\alpha^\prime}} \big] \times \\ C_{\ell} & (\alpha,\gamma)
		+ \big[\ket{E_{\gamma^\prime}} \bra{E_{\gamma^\prime}} \hat{\sigma}^\ell_+\ket{E_{\alpha^\prime}} \bra{E_{\alpha^\prime}}  ,  \\ & \ket{E_\alpha} \bra{E_\alpha} \hat{\sigma}^\ell_-  \ket{E_\gamma} \bra{E_\gamma} \hat{\rho} \big]  D_{\ell}(\alpha,\gamma) + \text{H.c}  \Big\} ,
		\label{ele}
	\end{aligned}
\end{equation}
where the $C_l$ and $D_l$ constants are the same as in Redfield, given by Eq.(\ref{redfield:constants}).

\section{Universal Lindblad equation}
\label{appendix:ULE}
The universal lindblad equation for our setup is derived from the steps described in \cite{ule}. 
The ULE equation is then given by
\begin{equation}
	\begin{aligned}
		\frac{\partial \hat{\rho}}{\partial t}=&-i[\hat{H}_S+\hat{H}_{LS},\hat{\rho}]\\+&\sum_{\ell,k} \Big( \hat{L}_{(\ell,k)} \hat{\rho} \hat{L}_{(\ell,k)}^\dagger - \frac{1}{2} \{ \hat{L}_{(\ell,k)}^\dagger \hat{L}_{(\ell,k)}, \hat{\rho}   \} \Big)
	\end{aligned}
\end{equation}
The lamb shift $\hat{H}_{LS}$ and Lindblad operators $\hat{L}_{(\ell,k)}$ are given by, 
\begin{equation}
	\begin{aligned}
		\hat{L}_{(\ell,k)}=&{2 \pi \epsilon} \sum_{\alpha,\gamma, \ell^\prime, k^\prime} \bm{g}_{(\ell,\ell^\prime,k,k^\prime)}(E_{\gamma\alpha})  \bm{X}^{(\ell^\prime,k^\prime)}_{\alpha \gamma}   \ket{E_\alpha} \bra{E_\gamma}\\
		\hat{H}_{LS}=& \hspace{-1em} \sum_{\alpha \gamma \eta;\ell k \ell^\prime k^\prime}  \hspace{-1em} \bm{X}^{(\ell,k)}_{\alpha \eta} \bm{X}^{(\ell^\prime,k^\prime)}_{\eta \gamma} \bm{f}_{\ell,\ell^\prime,k,k^\prime}(E_{\alpha \eta},E_{\gamma \eta})  \ket{E_\alpha} \bra{E_\gamma} 
	\end{aligned}
\end{equation}
where $\bm{X}^{(\ell,k)}_{\alpha\gamma} = \bra{E_\alpha} \hat{X}_{(\ell,k)} \ket{E_\gamma}$, and 
\begin{equation}
	\bm{f}(p,q)=-2 \pi \epsilon^2 \mathcal{P} \int_{-\infty}^{\infty} d\omega \frac{\bm{g}(\omega-p) \bm{g} (\omega+q)}{\omega}
\end{equation}
Note that $\bm{g}_{(\ell,\ell^\prime,k,k^\prime)}$ should be treated as a matrix with row index $(\ell,k)$ and column index $(\ell^\prime,k^\prime)$. Thus, $ [\bm{g}(\omega-p) \bm{g} (\omega+q)]_{(\ell,\ell^{\prime \prime} , k, k^{\prime \prime})}=\sum_{(\ell^\prime,k^\prime)} \bm{g}(\omega-p)_{(\ell,\ell^\prime,k,k^\prime)} \bm{g}(\omega+q)_{(\ell^\prime,\ell^{\prime \prime},k^\prime,k^{\prime \prime})}$. The $\bm{g}(\omega)$ matrix captures the effect of the bath on the system, and can be evaluated to be  
\begin{equation}
	\bm{g}_{(\ell,\ell^\prime,k,k^\prime)} (\omega)=\delta_{\ell,\ell^\prime} 
	\begin{cases*}
		\frac{\sqrt{\mathfrak{J}_\ell(-\omega) n_\ell(-\omega)}}{4 \sqrt{2} \pi}  \begin{pmatrix} 1 & i \\-i & 1 \end{pmatrix}_{k,k^\prime} & \hspace{-1.5em} if $\omega<0$ \\
		\frac{\sqrt{\mathfrak{J}_\ell(\omega) (1+n_\ell(\omega))}}{4 \sqrt{2} \pi} \begin{pmatrix} 1 & -i \\i & 1 \end{pmatrix}_{k,k^\prime} & \hspace{-1.5em} if $\omega>0$ 
	\end{cases*}
\end{equation} 
As in the Redfield case, $\ket{E_\alpha}$, $\ket{E_\gamma}$ are eigenkets of the system Hamiltonian with eigenenergies $E_\alpha$ and $E_\gamma$, and $E_{\gamma \alpha} = E_\gamma-E_\alpha$.

\bibliography{bibliography}

\begin{thebibliography}{66}%
\makeatletter
\providecommand \@ifxundefined [1]{%
 \@ifx{#1\undefined}
}%
\providecommand \@ifnum [1]{%
 \ifnum #1\expandafter \@firstoftwo
 \else \expandafter \@secondoftwo
 \fi
}%
\providecommand \@ifx [1]{%
 \ifx #1\expandafter \@firstoftwo
 \else \expandafter \@secondoftwo
 \fi
}%
\providecommand \natexlab [1]{#1}%
\providecommand \enquote  [1]{``#1''}%
\providecommand \bibnamefont  [1]{#1}%
\providecommand \bibfnamefont [1]{#1}%
\providecommand \citenamefont [1]{#1}%
\providecommand \href@noop [0]{\@secondoftwo}%
\providecommand \href [0]{\begingroup \@sanitize@url \@href}%
\providecommand \@href[1]{\@@startlink{#1}\@@href}%
\providecommand \@@href[1]{\endgroup#1\@@endlink}%
\providecommand \@sanitize@url [0]{\catcode `\\12\catcode `\$12\catcode
  `\&12\catcode `\#12\catcode `\^12\catcode `\_12\catcode `\%12\relax}%
\providecommand \@@startlink[1]{}%
\providecommand \@@endlink[0]{}%
\providecommand \url  [0]{\begingroup\@sanitize@url \@url }%
\providecommand \@url [1]{\endgroup\@href {#1}{\urlprefix }}%
\providecommand \urlprefix  [0]{URL }%
\providecommand \Eprint [0]{\href }%
\providecommand \doibase [0]{http://dx.doi.org/}%
\providecommand \selectlanguage [0]{\@gobble}%
\providecommand \bibinfo  [0]{\@secondoftwo}%
\providecommand \bibfield  [0]{\@secondoftwo}%
\providecommand \translation [1]{[#1]}%
\providecommand \BibitemOpen [0]{}%
\providecommand \bibitemStop [0]{}%
\providecommand \bibitemNoStop [0]{.\EOS\space}%
\providecommand \EOS [0]{\spacefactor3000\relax}%
\providecommand \BibitemShut  [1]{\csname bibitem#1\endcsname}%
\let\auto@bib@innerbib\@empty
\bibitem [{\citenamefont {Hur}\ \emph {et~al.}(2016)\citenamefont {Hur},
  \citenamefont {Henriet}, \citenamefont {Petrescu}, \citenamefont {Plekhanov},
  \citenamefont {Roux},\ and\ \citenamefont {Schir{\'{o}}}}]{LeHur2016}%
  \BibitemOpen
  \bibfield  {author} {\bibinfo {author} {\bibfnamefont {K.~L.}\ \bibnamefont
  {Hur}}, \bibinfo {author} {\bibfnamefont {L.}~\bibnamefont {Henriet}},
  \bibinfo {author} {\bibfnamefont {A.}~\bibnamefont {Petrescu}}, \bibinfo
  {author} {\bibfnamefont {K.}~\bibnamefont {Plekhanov}}, \bibinfo {author}
  {\bibfnamefont {G.}~\bibnamefont {Roux}}, \ and\ \bibinfo {author}
  {\bibfnamefont {M.}~\bibnamefont {Schir{\'{o}}}},\ }\href {\doibase
  10.1016/j.crhy.2016.05.003} {\bibfield  {journal} {\bibinfo  {journal}
  {Comptes Rendus Physique}\ }\textbf {\bibinfo {volume} {17}},\ \bibinfo
  {pages} {808} (\bibinfo {year} {2016})}\BibitemShut {NoStop}%
\bibitem [{\citenamefont {Binder}\ \emph {et~al.}(2018)\citenamefont {Binder},
  \citenamefont {Correa}, \citenamefont {Gogolin}, \citenamefont {Anders},\
  and\ \citenamefont {Adesso}}]{quantum_thermodynamics}%
  \BibitemOpen
  \bibinfo {editor} {\bibfnamefont {F.}~\bibnamefont {Binder}}, \bibinfo
  {editor} {\bibfnamefont {L.~A.}\ \bibnamefont {Correa}}, \bibinfo {editor}
  {\bibfnamefont {C.}~\bibnamefont {Gogolin}}, \bibinfo {editor} {\bibfnamefont
  {J.}~\bibnamefont {Anders}}, \ and\ \bibinfo {editor} {\bibfnamefont
  {G.}~\bibnamefont {Adesso}},\ eds.,\ \href@noop {} {\emph {\bibinfo {title}
  {Thermodynamics in the Quantum Regime}}}\ (\bibinfo  {publisher} {Springer,
  Cham},\ \bibinfo {year} {2018})\BibitemShut {NoStop}%
\bibitem [{\citenamefont {Cao}\ \emph {et~al.}(2019)\citenamefont {Cao},
  \citenamefont {Romero}, \citenamefont {Olson}, \citenamefont {Degroote},
  \citenamefont {Johnson}, \citenamefont {Kieferová}, \citenamefont
  {Kivlichan}, \citenamefont {Menke}, \citenamefont {Peropadre}, \citenamefont
  {Sawaya},\ and\ \citenamefont {et~al.}}]{quantum_chemistry}%
  \BibitemOpen
  \bibfield  {author} {\bibinfo {author} {\bibfnamefont {Y.}~\bibnamefont
  {Cao}}, \bibinfo {author} {\bibfnamefont {J.}~\bibnamefont {Romero}},
  \bibinfo {author} {\bibfnamefont {J.~P.}\ \bibnamefont {Olson}}, \bibinfo
  {author} {\bibfnamefont {M.}~\bibnamefont {Degroote}}, \bibinfo {author}
  {\bibfnamefont {P.~D.}\ \bibnamefont {Johnson}}, \bibinfo {author}
  {\bibfnamefont {M.}~\bibnamefont {Kieferová}}, \bibinfo {author}
  {\bibfnamefont {I.~D.}\ \bibnamefont {Kivlichan}}, \bibinfo {author}
  {\bibfnamefont {T.}~\bibnamefont {Menke}}, \bibinfo {author} {\bibfnamefont
  {B.}~\bibnamefont {Peropadre}}, \bibinfo {author} {\bibfnamefont {N.~P.~D.}\
  \bibnamefont {Sawaya}}, \ and\ \bibinfo {author} {\bibnamefont {et~al.}},\
  }\href {http://dx.doi.org/10.1021/acs.chemrev.8b00803} {\bibfield  {journal}
  {\bibinfo  {journal} {Chemical Reviews}\ }\textbf {\bibinfo {volume} {119}},\
  \bibinfo {pages} {10856–10915} (\bibinfo {year} {2019})}\BibitemShut
  {NoStop}%
\bibitem [{\citenamefont {Awschalom}\ \emph {et~al.}(2021)\citenamefont
  {Awschalom}, \citenamefont {Du}, \citenamefont {He}, \citenamefont
  {Heremans}, \citenamefont {Hoffmann}, \citenamefont {Hou}, \citenamefont
  {Kurebayashi}, \citenamefont {Li}, \citenamefont {Liu}, \citenamefont
  {Novosad}, \citenamefont {Sklenar}, \citenamefont {Sullivan}, \citenamefont
  {Sun}, \citenamefont {Tang}, \citenamefont {Tyberkevych}, \citenamefont
  {Trevillian}, \citenamefont {Tsen}, \citenamefont {Weiss}, \citenamefont
  {Zhang}, \citenamefont {Zhang}, \citenamefont {Zhao},\ and\ \citenamefont
  {Zollitsch}}]{quantum_engineering}%
  \BibitemOpen
  \bibfield  {author} {\bibinfo {author} {\bibfnamefont {D.~D.}\ \bibnamefont
  {Awschalom}}, \bibinfo {author} {\bibfnamefont {C.~H.~R.}\ \bibnamefont
  {Du}}, \bibinfo {author} {\bibfnamefont {R.}~\bibnamefont {He}}, \bibinfo
  {author} {\bibfnamefont {J.}~\bibnamefont {Heremans}}, \bibinfo {author}
  {\bibfnamefont {A.}~\bibnamefont {Hoffmann}}, \bibinfo {author}
  {\bibfnamefont {J.}~\bibnamefont {Hou}}, \bibinfo {author} {\bibfnamefont
  {H.}~\bibnamefont {Kurebayashi}}, \bibinfo {author} {\bibfnamefont
  {Y.}~\bibnamefont {Li}}, \bibinfo {author} {\bibfnamefont {L.}~\bibnamefont
  {Liu}}, \bibinfo {author} {\bibfnamefont {V.}~\bibnamefont {Novosad}},
  \bibinfo {author} {\bibfnamefont {J.}~\bibnamefont {Sklenar}}, \bibinfo
  {author} {\bibfnamefont {S.}~\bibnamefont {Sullivan}}, \bibinfo {author}
  {\bibfnamefont {D.}~\bibnamefont {Sun}}, \bibinfo {author} {\bibfnamefont
  {H.}~\bibnamefont {Tang}}, \bibinfo {author} {\bibfnamefont {V.}~\bibnamefont
  {Tyberkevych}}, \bibinfo {author} {\bibfnamefont {C.}~\bibnamefont
  {Trevillian}}, \bibinfo {author} {\bibfnamefont {A.~W.}\ \bibnamefont
  {Tsen}}, \bibinfo {author} {\bibfnamefont {L.}~\bibnamefont {Weiss}},
  \bibinfo {author} {\bibfnamefont {W.}~\bibnamefont {Zhang}}, \bibinfo
  {author} {\bibfnamefont {X.}~\bibnamefont {Zhang}}, \bibinfo {author}
  {\bibfnamefont {L.}~\bibnamefont {Zhao}}, \ and\ \bibinfo {author}
  {\bibfnamefont {C.~W.}\ \bibnamefont {Zollitsch}},\ }\href {\doibase
  10.1109/TQE.2021.3057799} {\bibfield  {journal} {\bibinfo  {journal} {IEEE
  Transactions on Quantum Engineering}\ ,\ \bibinfo {pages} {1}} (\bibinfo
  {year} {2021})}\BibitemShut {NoStop}%
\bibitem [{\citenamefont {Cao}\ \emph {et~al.}(2020)\citenamefont {Cao},
  \citenamefont {Cogdell}, \citenamefont {Coker}, \citenamefont {Duan},
  \citenamefont {Hauer}, \citenamefont {Kleinekath{\"o}fer}, \citenamefont
  {Jansen}, \citenamefont {Man{\v c}al}, \citenamefont {Miller}, \citenamefont
  {Ogilvie}, \citenamefont {Prokhorenko}, \citenamefont {Renger}, \citenamefont
  {Tan}, \citenamefont {Tempelaar}, \citenamefont {Thorwart}, \citenamefont
  {Thyrhaug}, \citenamefont {Westenhoff},\ and\ \citenamefont
  {Zigmantas}}]{quantum_biology}%
  \BibitemOpen
  \bibfield  {author} {\bibinfo {author} {\bibfnamefont {J.}~\bibnamefont
  {Cao}}, \bibinfo {author} {\bibfnamefont {R.~J.}\ \bibnamefont {Cogdell}},
  \bibinfo {author} {\bibfnamefont {D.~F.}\ \bibnamefont {Coker}}, \bibinfo
  {author} {\bibfnamefont {H.-G.}\ \bibnamefont {Duan}}, \bibinfo {author}
  {\bibfnamefont {J.}~\bibnamefont {Hauer}}, \bibinfo {author} {\bibfnamefont
  {U.}~\bibnamefont {Kleinekath{\"o}fer}}, \bibinfo {author} {\bibfnamefont
  {T.~L.~C.}\ \bibnamefont {Jansen}}, \bibinfo {author} {\bibfnamefont
  {T.}~\bibnamefont {Man{\v c}al}}, \bibinfo {author} {\bibfnamefont
  {R.~J.~D.}\ \bibnamefont {Miller}}, \bibinfo {author} {\bibfnamefont {J.~P.}\
  \bibnamefont {Ogilvie}}, \bibinfo {author} {\bibfnamefont {V.~I.}\
  \bibnamefont {Prokhorenko}}, \bibinfo {author} {\bibfnamefont
  {T.}~\bibnamefont {Renger}}, \bibinfo {author} {\bibfnamefont {H.-S.}\
  \bibnamefont {Tan}}, \bibinfo {author} {\bibfnamefont {R.}~\bibnamefont
  {Tempelaar}}, \bibinfo {author} {\bibfnamefont {M.}~\bibnamefont {Thorwart}},
  \bibinfo {author} {\bibfnamefont {E.}~\bibnamefont {Thyrhaug}}, \bibinfo
  {author} {\bibfnamefont {S.}~\bibnamefont {Westenhoff}}, \ and\ \bibinfo
  {author} {\bibfnamefont {D.}~\bibnamefont {Zigmantas}},\ }\href
  {https://advances.sciencemag.org/content/6/14/eaaz4888} {\bibfield  {journal}
  {\bibinfo  {journal} {Science Advances}\ }\textbf {\bibinfo {volume} {6}}
  (\bibinfo {year} {2020})}\BibitemShut {NoStop}%
\bibitem [{\citenamefont {Gorini}\ \emph {et~al.}(1976)\citenamefont {Gorini},
  \citenamefont {Kossakowski},\ and\ \citenamefont {Sudarshan}}]{GKS1976}%
  \BibitemOpen
  \bibfield  {author} {\bibinfo {author} {\bibfnamefont {V.}~\bibnamefont
  {Gorini}}, \bibinfo {author} {\bibfnamefont {A.}~\bibnamefont {Kossakowski}},
  \ and\ \bibinfo {author} {\bibfnamefont {E.~C.~G.}\ \bibnamefont
  {Sudarshan}},\ }\href {\doibase 10.1063/1.522979} {\bibfield  {journal}
  {\bibinfo  {journal} {Journal of Mathematical Physics}\ }\textbf {\bibinfo
  {volume} {17}},\ \bibinfo {pages} {821} (\bibinfo {year} {1976})}\BibitemShut
  {NoStop}%
\bibitem [{\citenamefont {Lindblad}(1976)}]{lindblad1976}%
  \BibitemOpen
  \bibfield  {author} {\bibinfo {author} {\bibfnamefont {G.}~\bibnamefont
  {Lindblad}},\ }\href {\doibase 10.1007/bf01608499} {\bibfield  {journal}
  {\bibinfo  {journal} {Communications in Mathematical Physics}\ }\textbf
  {\bibinfo {volume} {48}},\ \bibinfo {pages} {119} (\bibinfo {year}
  {1976})}\BibitemShut {NoStop}%
\bibitem [{\citenamefont {Breuer}\ and\ \citenamefont
  {Petruccione}(2006)}]{breuer_book}%
  \BibitemOpen
  \bibfield  {author} {\bibinfo {author} {\bibfnamefont {H.-P.}\ \bibnamefont
  {Breuer}}\ and\ \bibinfo {author} {\bibfnamefont {F.}~\bibnamefont
  {Petruccione}},\ }\href@noop {} {\emph {\bibinfo {title} {The Theory of Open
  Quantum Systems}}}\ (\bibinfo  {publisher} {Oxford University Press,
  Oxford},\ \bibinfo {year} {2006})\BibitemShut {NoStop}%
\bibitem [{\citenamefont {Carmichael}(2002)}]{carmichael_book}%
  \BibitemOpen
  \bibfield  {author} {\bibinfo {author} {\bibfnamefont {H.}~\bibnamefont
  {Carmichael}},\ }\href@noop {} {\emph {\bibinfo {title} {Statistical Methods
  in Quantum Optics 1. Master Equations and Fokker–Planck Equations}}}\
  (\bibinfo  {publisher} {Springer-Verlag Berlin Heidelberg},\ \bibinfo {year}
  {2002})\BibitemShut {NoStop}%
\bibitem [{\citenamefont {Rivas}\ and\ \citenamefont
  {Huelga}(2012)}]{Rivas_2012}%
  \BibitemOpen
  \bibfield  {author} {\bibinfo {author} {\bibfnamefont {{\'A}.}~\bibnamefont
  {Rivas}}\ and\ \bibinfo {author} {\bibfnamefont {S.~F.}\ \bibnamefont
  {Huelga}},\ }\href {http://dx.doi.org/10.1007/978-3-642-23354-8} {\bibfield
  {journal} {\bibinfo  {journal} {SpringerBriefs in Physics}\ } (\bibinfo
  {year} {2012})}\BibitemShut {NoStop}%
\bibitem [{\citenamefont {Weimer}\ \emph {et~al.}(2021)\citenamefont {Weimer},
  \citenamefont {Kshetrimayum},\ and\ \citenamefont {Or\'us}}]{Weimer_2021}%
  \BibitemOpen
  \bibfield  {author} {\bibinfo {author} {\bibfnamefont {H.}~\bibnamefont
  {Weimer}}, \bibinfo {author} {\bibfnamefont {A.}~\bibnamefont
  {Kshetrimayum}}, \ and\ \bibinfo {author} {\bibfnamefont {R.}~\bibnamefont
  {Or\'us}},\ }\href {\doibase 10.1103/RevModPhys.93.015008} {\bibfield
  {journal} {\bibinfo  {journal} {Rev. Mod. Phys.}\ }\textbf {\bibinfo {volume}
  {93}},\ \bibinfo {pages} {015008} (\bibinfo {year} {2021})}\BibitemShut
  {NoStop}%
\bibitem [{\citenamefont {Plenio}\ and\ \citenamefont
  {Knight}(1998)}]{Plenio_1998}%
  \BibitemOpen
  \bibfield  {author} {\bibinfo {author} {\bibfnamefont {M.~B.}\ \bibnamefont
  {Plenio}}\ and\ \bibinfo {author} {\bibfnamefont {P.~L.}\ \bibnamefont
  {Knight}},\ }\href {\doibase 10.1103/RevModPhys.70.101} {\bibfield  {journal}
  {\bibinfo  {journal} {Rev. Mod. Phys.}\ }\textbf {\bibinfo {volume} {70}},\
  \bibinfo {pages} {101} (\bibinfo {year} {1998})}\BibitemShut {NoStop}%
\bibitem [{\citenamefont {Dalibard}\ \emph {et~al.}(1992)\citenamefont
  {Dalibard}, \citenamefont {Castin},\ and\ \citenamefont
  {M{\o}lmer}}]{Dalibard1992}%
  \BibitemOpen
  \bibfield  {author} {\bibinfo {author} {\bibfnamefont {J.}~\bibnamefont
  {Dalibard}}, \bibinfo {author} {\bibfnamefont {Y.}~\bibnamefont {Castin}}, \
  and\ \bibinfo {author} {\bibfnamefont {K.}~\bibnamefont {M{\o}lmer}},\ }\href
  {\doibase 10.1103/physrevlett.68.580} {\bibfield  {journal} {\bibinfo
  {journal} {Phys. Rev. Lett.}\ }\textbf {\bibinfo {volume} {68}},\ \bibinfo
  {pages} {580} (\bibinfo {year} {1992})}\BibitemShut {NoStop}%
\bibitem [{\citenamefont {M{\o}lmer}\ \emph {et~al.}(1993)\citenamefont
  {M{\o}lmer}, \citenamefont {Castin},\ and\ \citenamefont
  {Dalibard}}]{Mlmer1993}%
  \BibitemOpen
  \bibfield  {author} {\bibinfo {author} {\bibfnamefont {K.}~\bibnamefont
  {M{\o}lmer}}, \bibinfo {author} {\bibfnamefont {Y.}~\bibnamefont {Castin}}, \
  and\ \bibinfo {author} {\bibfnamefont {J.}~\bibnamefont {Dalibard}},\ }\href
  {\doibase 10.1364/josab.10.000524} {\bibfield  {journal} {\bibinfo  {journal}
  {Journal of the Optical Society of America B}\ }\textbf {\bibinfo {volume}
  {10}},\ \bibinfo {pages} {524} (\bibinfo {year} {1993})}\BibitemShut
  {NoStop}%
\bibitem [{\citenamefont {Redfield}(1965)}]{redfield1965}%
  \BibitemOpen
  \bibfield  {author} {\bibinfo {author} {\bibfnamefont {A.}~\bibnamefont
  {Redfield}},\ }in\ \href {\doibase 10.1016/b978-1-4832-3114-3.50007-6} {\emph
  {\bibinfo {booktitle} {Advances in Magnetic Resonance}}}\ (\bibinfo
  {publisher} {Elsevier},\ \bibinfo {year} {1965})\ pp.\ \bibinfo {pages}
  {1--32}\BibitemShut {NoStop}%
\bibitem [{\citenamefont {Hartmann}\ and\ \citenamefont
  {Strunz}(2020)}]{Hartmann_2020_1}%
  \BibitemOpen
  \bibfield  {author} {\bibinfo {author} {\bibfnamefont {R.}~\bibnamefont
  {Hartmann}}\ and\ \bibinfo {author} {\bibfnamefont {W.~T.}\ \bibnamefont
  {Strunz}},\ }\href {\doibase 10.1103/PhysRevA.101.012103} {\bibfield
  {journal} {\bibinfo  {journal} {Phys. Rev. A}\ }\textbf {\bibinfo {volume}
  {101}},\ \bibinfo {pages} {012103} (\bibinfo {year} {2020})}\BibitemShut
  {NoStop}%
\bibitem [{\citenamefont {Eastham}\ \emph {et~al.}(2016)\citenamefont
  {Eastham}, \citenamefont {Kirton}, \citenamefont {Cammack}, \citenamefont
  {Lovett},\ and\ \citenamefont {Keeling}}]{Eastham_2016}%
  \BibitemOpen
  \bibfield  {author} {\bibinfo {author} {\bibfnamefont {P.~R.}\ \bibnamefont
  {Eastham}}, \bibinfo {author} {\bibfnamefont {P.}~\bibnamefont {Kirton}},
  \bibinfo {author} {\bibfnamefont {H.~M.}\ \bibnamefont {Cammack}}, \bibinfo
  {author} {\bibfnamefont {B.~W.}\ \bibnamefont {Lovett}}, \ and\ \bibinfo
  {author} {\bibfnamefont {J.}~\bibnamefont {Keeling}},\ }\href {\doibase
  10.1103/PhysRevA.94.012110} {\bibfield  {journal} {\bibinfo  {journal} {Phys.
  Rev. A}\ }\textbf {\bibinfo {volume} {94}},\ \bibinfo {pages} {012110}
  (\bibinfo {year} {2016})}\BibitemShut {NoStop}%
\bibitem [{\citenamefont {Anderloni}\ \emph {et~al.}(2007)\citenamefont
  {Anderloni}, \citenamefont {Benatti},\ and\ \citenamefont
  {Floreanini}}]{anderloni_2007}%
  \BibitemOpen
  \bibfield  {author} {\bibinfo {author} {\bibfnamefont {S.}~\bibnamefont
  {Anderloni}}, \bibinfo {author} {\bibfnamefont {F.}~\bibnamefont {Benatti}},
  \ and\ \bibinfo {author} {\bibfnamefont {R.}~\bibnamefont {Floreanini}},\
  }\href {\doibase 10.1088/1751-8113/40/7/013} {\bibfield  {journal} {\bibinfo
  {journal} {Journal of Physics A: Mathematical and Theoretical}\ }\textbf
  {\bibinfo {volume} {40}},\ \bibinfo {pages} {1625} (\bibinfo {year}
  {2007})}\BibitemShut {NoStop}%
\bibitem [{\citenamefont {Gaspard}\ and\ \citenamefont
  {Nagaoka}(1999)}]{Gaspard_Nagaoka_1999}%
  \BibitemOpen
  \bibfield  {author} {\bibinfo {author} {\bibfnamefont {P.}~\bibnamefont
  {Gaspard}}\ and\ \bibinfo {author} {\bibfnamefont {M.}~\bibnamefont
  {Nagaoka}},\ }\href {\doibase 10.1063/1.479867} {\bibfield  {journal}
  {\bibinfo  {journal} {The Journal of Chemical Physics}\ }\textbf {\bibinfo
  {volume} {111}},\ \bibinfo {pages} {5668} (\bibinfo {year}
  {1999})}\BibitemShut {NoStop}%
\bibitem [{\citenamefont {Kohen}\ \emph {et~al.}(1997)\citenamefont {Kohen},
  \citenamefont {Marston},\ and\ \citenamefont {Tannor}}]{Kohen_1997}%
  \BibitemOpen
  \bibfield  {author} {\bibinfo {author} {\bibfnamefont {D.}~\bibnamefont
  {Kohen}}, \bibinfo {author} {\bibfnamefont {C.~C.}\ \bibnamefont {Marston}},
  \ and\ \bibinfo {author} {\bibfnamefont {D.~J.}\ \bibnamefont {Tannor}},\
  }\href {\doibase 10.1063/1.474887} {\bibfield  {journal} {\bibinfo  {journal}
  {The Journal of Chemical Physics}\ }\textbf {\bibinfo {volume} {107}},\
  \bibinfo {pages} {5236} (\bibinfo {year} {1997})}\BibitemShut {NoStop}%
\bibitem [{\citenamefont {Gnutzmann}\ and\ \citenamefont
  {Haake}(1996)}]{Gnutzmann_1996}%
  \BibitemOpen
  \bibfield  {author} {\bibinfo {author} {\bibfnamefont {S.}~\bibnamefont
  {Gnutzmann}}\ and\ \bibinfo {author} {\bibfnamefont {F.}~\bibnamefont
  {Haake}},\ }\href {\doibase 10.1007/s002570050208} {\bibfield  {journal}
  {\bibinfo  {journal} {Zeitschrift f{\"u}r Physik B Condensed Matter}\
  }\textbf {\bibinfo {volume} {101}},\ \bibinfo {pages} {263} (\bibinfo {year}
  {1996})}\BibitemShut {NoStop}%
\bibitem [{\citenamefont {Suárez}\ \emph {et~al.}(1992)\citenamefont
  {Suárez}, \citenamefont {Silbey},\ and\ \citenamefont
  {Oppenheim}}]{Suarez_1992}%
  \BibitemOpen
  \bibfield  {author} {\bibinfo {author} {\bibfnamefont {A.}~\bibnamefont
  {Suárez}}, \bibinfo {author} {\bibfnamefont {R.}~\bibnamefont {Silbey}}, \
  and\ \bibinfo {author} {\bibfnamefont {I.}~\bibnamefont {Oppenheim}},\ }\href
  {\doibase 10.1063/1.463831} {\bibfield  {journal} {\bibinfo  {journal} {The
  Journal of Chemical Physics}\ }\textbf {\bibinfo {volume} {97}},\ \bibinfo
  {pages} {5101} (\bibinfo {year} {1992})}\BibitemShut {NoStop}%
\bibitem [{\citenamefont {Walls}(1970)}]{Walls1970}%
  \BibitemOpen
  \bibfield  {author} {\bibinfo {author} {\bibfnamefont {D.~F.}\ \bibnamefont
  {Walls}},\ }\href {\doibase 10.1007/bf01396784} {\bibfield  {journal}
  {\bibinfo  {journal} {Zeitschrift f\"{u}r Physik A Hadrons and nuclei}\
  }\textbf {\bibinfo {volume} {234}},\ \bibinfo {pages} {231} (\bibinfo {year}
  {1970})}\BibitemShut {NoStop}%
\bibitem [{\citenamefont {Wichterich}\ \emph {et~al.}(2007)\citenamefont
  {Wichterich}, \citenamefont {Henrich}, \citenamefont {Breuer}, \citenamefont
  {Gemmer},\ and\ \citenamefont {Michel}}]{Wichterich_2007}%
  \BibitemOpen
  \bibfield  {author} {\bibinfo {author} {\bibfnamefont {H.}~\bibnamefont
  {Wichterich}}, \bibinfo {author} {\bibfnamefont {M.~J.}\ \bibnamefont
  {Henrich}}, \bibinfo {author} {\bibfnamefont {H.-P.}\ \bibnamefont {Breuer}},
  \bibinfo {author} {\bibfnamefont {J.}~\bibnamefont {Gemmer}}, \ and\ \bibinfo
  {author} {\bibfnamefont {M.}~\bibnamefont {Michel}},\ }\href {\doibase
  10.1103/PhysRevE.76.031115} {\bibfield  {journal} {\bibinfo  {journal} {Phys.
  Rev. E}\ }\textbf {\bibinfo {volume} {76}},\ \bibinfo {pages} {031115}
  (\bibinfo {year} {2007})}\BibitemShut {NoStop}%
\bibitem [{\citenamefont {Rivas}\ \emph {et~al.}(2010)\citenamefont {Rivas},
  \citenamefont {Plato}, \citenamefont {Huelga},\ and\ \citenamefont
  {Plenio}}]{Rivas_2010}%
  \BibitemOpen
  \bibfield  {author} {\bibinfo {author} {\bibfnamefont {{\'{A}}.}~\bibnamefont
  {Rivas}}, \bibinfo {author} {\bibfnamefont {A.~D.~K.}\ \bibnamefont {Plato}},
  \bibinfo {author} {\bibfnamefont {S.~F.}\ \bibnamefont {Huelga}}, \ and\
  \bibinfo {author} {\bibfnamefont {M.~B.}\ \bibnamefont {Plenio}},\ }\href
  {\doibase 10.1088/1367-2630/12/11/113032} {\bibfield  {journal} {\bibinfo
  {journal} {New Journal of Physics}\ }\textbf {\bibinfo {volume} {12}},\
  \bibinfo {pages} {113032} (\bibinfo {year} {2010})}\BibitemShut {NoStop}%
\bibitem [{\citenamefont {Deçordi}\ and\ \citenamefont
  {Vidiella-Barranco}(2017)}]{barranco_2014}%
  \BibitemOpen
  \bibfield  {author} {\bibinfo {author} {\bibfnamefont {G.}~\bibnamefont
  {Deçordi}}\ and\ \bibinfo {author} {\bibfnamefont {A.}~\bibnamefont
  {Vidiella-Barranco}},\ }\href {\doibase
  https://doi.org/10.1016/j.optcom.2016.10.017} {\bibfield  {journal} {\bibinfo
   {journal} {Optics Communications}\ }\textbf {\bibinfo {volume} {387}},\
  \bibinfo {pages} {366} (\bibinfo {year} {2017})}\BibitemShut {NoStop}%
\bibitem [{\citenamefont {Levy}\ and\ \citenamefont
  {Kosloff}(2014)}]{Levy2014}%
  \BibitemOpen
  \bibfield  {author} {\bibinfo {author} {\bibfnamefont {A.}~\bibnamefont
  {Levy}}\ and\ \bibinfo {author} {\bibfnamefont {R.}~\bibnamefont {Kosloff}},\
  }\href {\doibase 10.1209/0295-5075/107/20004} {\bibfield  {journal} {\bibinfo
   {journal} {{EPL} (Europhysics Letters)}\ }\textbf {\bibinfo {volume}
  {107}},\ \bibinfo {pages} {20004} (\bibinfo {year} {2014})}\BibitemShut
  {NoStop}%
\bibitem [{\citenamefont {Purkayastha}\ \emph {et~al.}(2016)\citenamefont
  {Purkayastha}, \citenamefont {Dhar},\ and\ \citenamefont
  {Kulkarni}}]{archak}%
  \BibitemOpen
  \bibfield  {author} {\bibinfo {author} {\bibfnamefont {A.}~\bibnamefont
  {Purkayastha}}, \bibinfo {author} {\bibfnamefont {A.}~\bibnamefont {Dhar}}, \
  and\ \bibinfo {author} {\bibfnamefont {M.}~\bibnamefont {Kulkarni}},\ }\href
  {\doibase 10.1103/PhysRevA.93.062114} {\bibfield  {journal} {\bibinfo
  {journal} {Phys. Rev. A}\ }\textbf {\bibinfo {volume} {93}},\ \bibinfo
  {pages} {062114} (\bibinfo {year} {2016})}\BibitemShut {NoStop}%
\bibitem [{\citenamefont {Trushechkin}\ and\ \citenamefont
  {Volovich}(2016)}]{Trushechkin_2016}%
  \BibitemOpen
  \bibfield  {author} {\bibinfo {author} {\bibfnamefont {A.~S.}\ \bibnamefont
  {Trushechkin}}\ and\ \bibinfo {author} {\bibfnamefont {I.~V.}\ \bibnamefont
  {Volovich}},\ }\href {\doibase 10.1209/0295-5075/113/30005} {\bibfield
  {journal} {\bibinfo  {journal} {{EPL} (Europhysics Letters)}\ }\textbf
  {\bibinfo {volume} {113}},\ \bibinfo {pages} {30005} (\bibinfo {year}
  {2016})}\BibitemShut {NoStop}%
\bibitem [{\citenamefont {Hofer}\ \emph {et~al.}(2017)\citenamefont {Hofer},
  \citenamefont {Perarnau-Llobet}, \citenamefont {Miranda}, \citenamefont
  {Haack}, \citenamefont {Silva}, \citenamefont {Brask},\ and\ \citenamefont
  {Brunner}}]{Hofer_2017}%
  \BibitemOpen
  \bibfield  {author} {\bibinfo {author} {\bibfnamefont {P.~P.}\ \bibnamefont
  {Hofer}}, \bibinfo {author} {\bibfnamefont {M.}~\bibnamefont
  {Perarnau-Llobet}}, \bibinfo {author} {\bibfnamefont {L.~D.~M.}\ \bibnamefont
  {Miranda}}, \bibinfo {author} {\bibfnamefont {G.}~\bibnamefont {Haack}},
  \bibinfo {author} {\bibfnamefont {R.}~\bibnamefont {Silva}}, \bibinfo
  {author} {\bibfnamefont {J.~B.}\ \bibnamefont {Brask}}, \ and\ \bibinfo
  {author} {\bibfnamefont {N.}~\bibnamefont {Brunner}},\ }\href {\doibase
  10.1088/1367-2630/aa964f} {\bibfield  {journal} {\bibinfo  {journal} {New
  Journal of Physics}\ }\textbf {\bibinfo {volume} {19}},\ \bibinfo {pages}
  {123037} (\bibinfo {year} {2017})}\BibitemShut {NoStop}%
\bibitem [{\citenamefont {González}\ \emph {et~al.}(2017)\citenamefont
  {González}, \citenamefont {Correa}, \citenamefont {Nocerino}, \citenamefont
  {Palao}, \citenamefont {Alonso},\ and\ \citenamefont
  {Adesso}}]{Gonzalez_2017}%
  \BibitemOpen
  \bibfield  {author} {\bibinfo {author} {\bibfnamefont {J.~O.}\ \bibnamefont
  {González}}, \bibinfo {author} {\bibfnamefont {L.~A.}\ \bibnamefont
  {Correa}}, \bibinfo {author} {\bibfnamefont {G.}~\bibnamefont {Nocerino}},
  \bibinfo {author} {\bibfnamefont {J.~P.}\ \bibnamefont {Palao}}, \bibinfo
  {author} {\bibfnamefont {D.}~\bibnamefont {Alonso}}, \ and\ \bibinfo {author}
  {\bibfnamefont {G.}~\bibnamefont {Adesso}},\ }\href {\doibase
  10.1142/S1230161217400108} {\bibfield  {journal} {\bibinfo  {journal} {Open
  Systems \& Information Dynamics}\ }\textbf {\bibinfo {volume} {24}},\
  \bibinfo {pages} {1740010} (\bibinfo {year} {2017})}\BibitemShut {NoStop}%
\bibitem [{\citenamefont {Mitchison}\ and\ \citenamefont
  {Plenio}(2018)}]{Mitchison_2018}%
  \BibitemOpen
  \bibfield  {author} {\bibinfo {author} {\bibfnamefont {M.~T.}\ \bibnamefont
  {Mitchison}}\ and\ \bibinfo {author} {\bibfnamefont {M.~B.}\ \bibnamefont
  {Plenio}},\ }\href {\doibase 10.1088/1367-2630/aa9f70} {\bibfield  {journal}
  {\bibinfo  {journal} {New Journal of Physics}\ }\textbf {\bibinfo {volume}
  {20}},\ \bibinfo {pages} {033005} (\bibinfo {year} {2018})}\BibitemShut
  {NoStop}%
\bibitem [{\citenamefont {Cattaneo}\ \emph {et~al.}(2019)\citenamefont
  {Cattaneo}, \citenamefont {Giorgi}, \citenamefont {Maniscalco},\ and\
  \citenamefont {Zambrini}}]{Cattaneo_2019}%
  \BibitemOpen
  \bibfield  {author} {\bibinfo {author} {\bibfnamefont {M.}~\bibnamefont
  {Cattaneo}}, \bibinfo {author} {\bibfnamefont {G.~L.}\ \bibnamefont
  {Giorgi}}, \bibinfo {author} {\bibfnamefont {S.}~\bibnamefont {Maniscalco}},
  \ and\ \bibinfo {author} {\bibfnamefont {R.}~\bibnamefont {Zambrini}},\
  }\href {\doibase 10.1088/1367-2630/ab54ac} {\bibfield  {journal} {\bibinfo
  {journal} {New Journal of Physics}\ }\textbf {\bibinfo {volume} {21}},\
  \bibinfo {pages} {113045} (\bibinfo {year} {2019})}\BibitemShut {NoStop}%
\bibitem [{\citenamefont {Benatti}\ \emph {et~al.}(2020)\citenamefont
  {Benatti}, \citenamefont {Floreanini},\ and\ \citenamefont
  {Memarzadeh}}]{Benatti_2020}%
  \BibitemOpen
  \bibfield  {author} {\bibinfo {author} {\bibfnamefont {F.}~\bibnamefont
  {Benatti}}, \bibinfo {author} {\bibfnamefont {R.}~\bibnamefont {Floreanini}},
  \ and\ \bibinfo {author} {\bibfnamefont {L.}~\bibnamefont {Memarzadeh}},\
  }\href {\doibase 10.1103/PhysRevA.102.042219} {\bibfield  {journal} {\bibinfo
   {journal} {Phys. Rev. A}\ }\textbf {\bibinfo {volume} {102}},\ \bibinfo
  {pages} {042219} (\bibinfo {year} {2020})}\BibitemShut {NoStop}%
\bibitem [{\citenamefont {Konopik}\ and\ \citenamefont
  {Lutz}(2020)}]{konopik_2020local}%
  \BibitemOpen
  \bibfield  {author} {\bibinfo {author} {\bibfnamefont {M.}~\bibnamefont
  {Konopik}}\ and\ \bibinfo {author} {\bibfnamefont {E.}~\bibnamefont {Lutz}},\
  }\href@noop {} {\  (\bibinfo {year} {2020})},\ \Eprint
  {http://arxiv.org/abs/2012.09907} {arXiv:2012.09907 [quant-ph]} \BibitemShut
  {NoStop}%
\bibitem [{\citenamefont {Scali}\ \emph {et~al.}(2021)\citenamefont {Scali},
  \citenamefont {Anders},\ and\ \citenamefont {Correa}}]{Scali_2021}%
  \BibitemOpen
  \bibfield  {author} {\bibinfo {author} {\bibfnamefont {S.}~\bibnamefont
  {Scali}}, \bibinfo {author} {\bibfnamefont {J.}~\bibnamefont {Anders}}, \
  and\ \bibinfo {author} {\bibfnamefont {L.~A.}\ \bibnamefont {Correa}},\
  }\href {http://dx.doi.org/10.22331/q-2021-05-01-451} {\bibfield  {journal}
  {\bibinfo  {journal} {Quantum}\ }\textbf {\bibinfo {volume} {5}},\ \bibinfo
  {pages} {451} (\bibinfo {year} {2021})}\BibitemShut {NoStop}%
\bibitem [{\citenamefont {Benatti}\ \emph {et~al.}(2021)\citenamefont
  {Benatti}, \citenamefont {Floreanini},\ and\ \citenamefont
  {Memarzadeh}}]{Floreanini_2021}%
  \BibitemOpen
  \bibfield  {author} {\bibinfo {author} {\bibfnamefont {F.}~\bibnamefont
  {Benatti}}, \bibinfo {author} {\bibfnamefont {R.}~\bibnamefont {Floreanini}},
  \ and\ \bibinfo {author} {\bibfnamefont {L.}~\bibnamefont {Memarzadeh}},\
  }\href@noop {} {\  (\bibinfo {year} {2021})},\ \Eprint
  {http://arxiv.org/abs/2102.10036} {arXiv:2102.10036 [quant-ph]} \BibitemShut
  {NoStop}%
\bibitem [{\citenamefont {Trushechkin}(2021)}]{trushechkin2021}%
  \BibitemOpen
  \bibfield  {author} {\bibinfo {author} {\bibfnamefont {A.}~\bibnamefont
  {Trushechkin}},\ }\href {\doibase 10.1103/PhysRevA.103.062226} {\bibfield
  {journal} {\bibinfo  {journal} {Phys. Rev. A}\ }\textbf {\bibinfo {volume}
  {103}},\ \bibinfo {pages} {062226} (\bibinfo {year} {2021})}\BibitemShut
  {NoStop}%
\bibitem [{\citenamefont {Davidovic}(2021)}]{Davidovic_2021}%
  \BibitemOpen
  \bibfield  {author} {\bibinfo {author} {\bibfnamefont {D.}~\bibnamefont
  {Davidovic}},\ }\href {https://arxiv.org/abs/2112.07863} {\  (\bibinfo {year}
  {2021})},\ \Eprint {http://arxiv.org/abs/2112.07863} {arXiv:2112.07863
  [quant-ph]} \BibitemShut {NoStop}%
\bibitem [{\citenamefont {Nathan}\ and\ \citenamefont {Rudner}(2020)}]{ule}%
  \BibitemOpen
  \bibfield  {author} {\bibinfo {author} {\bibfnamefont {F.}~\bibnamefont
  {Nathan}}\ and\ \bibinfo {author} {\bibfnamefont {M.~S.}\ \bibnamefont
  {Rudner}},\ }\href {\doibase 10.1103/PhysRevB.102.115109} {\bibfield
  {journal} {\bibinfo  {journal} {Phys. Rev. B}\ }\textbf {\bibinfo {volume}
  {102}},\ \bibinfo {pages} {115109} (\bibinfo {year} {2020})}\BibitemShut
  {NoStop}%
\bibitem [{\citenamefont {Kleinherbers}\ \emph {et~al.}(2020)\citenamefont
  {Kleinherbers}, \citenamefont {Szpak}, \citenamefont {K\"onig},\ and\
  \citenamefont {Sch\"utzhold}}]{Kleinherbers_2020}%
  \BibitemOpen
  \bibfield  {author} {\bibinfo {author} {\bibfnamefont {E.}~\bibnamefont
  {Kleinherbers}}, \bibinfo {author} {\bibfnamefont {N.}~\bibnamefont {Szpak}},
  \bibinfo {author} {\bibfnamefont {J.}~\bibnamefont {K\"onig}}, \ and\
  \bibinfo {author} {\bibfnamefont {R.}~\bibnamefont {Sch\"utzhold}},\ }\href
  {\doibase 10.1103/PhysRevB.101.125131} {\bibfield  {journal} {\bibinfo
  {journal} {Phys. Rev. B}\ }\textbf {\bibinfo {volume} {101}},\ \bibinfo
  {pages} {125131} (\bibinfo {year} {2020})}\BibitemShut {NoStop}%
\bibitem [{\citenamefont {Davidovi{\'c}}(2020)}]{Davidovic_2020}%
  \BibitemOpen
  \bibfield  {author} {\bibinfo {author} {\bibfnamefont {D.}~\bibnamefont
  {Davidovi{\'c}}},\ }\href {\doibase 10.22331/q-2020-09-21-326} {\bibfield
  {journal} {\bibinfo  {journal} {Quantum}\ }\textbf {\bibinfo {volume} {4}},\
  \bibinfo {pages} {326} (\bibinfo {year} {2020})}\BibitemShut {NoStop}%
\bibitem [{\citenamefont {Mozgunov}\ and\ \citenamefont
  {Lidar}(2020)}]{mozgunov2020}%
  \BibitemOpen
  \bibfield  {author} {\bibinfo {author} {\bibfnamefont {E.}~\bibnamefont
  {Mozgunov}}\ and\ \bibinfo {author} {\bibfnamefont {D.}~\bibnamefont
  {Lidar}},\ }\href {http://dx.doi.org/10.22331/q-2020-02-06-227} {\bibfield
  {journal} {\bibinfo  {journal} {Quantum}\ }\textbf {\bibinfo {volume} {4}},\
  \bibinfo {pages} {227} (\bibinfo {year} {2020})}\BibitemShut {NoStop}%
\bibitem [{\citenamefont {McCauley}\ \emph {et~al.}(2020)\citenamefont
  {McCauley}, \citenamefont {Cruikshank}, \citenamefont {Bondar},\ and\
  \citenamefont {Jacobs}}]{mccauley2020}%
  \BibitemOpen
  \bibfield  {author} {\bibinfo {author} {\bibfnamefont {G.}~\bibnamefont
  {McCauley}}, \bibinfo {author} {\bibfnamefont {B.}~\bibnamefont
  {Cruikshank}}, \bibinfo {author} {\bibfnamefont {D.~I.}\ \bibnamefont
  {Bondar}}, \ and\ \bibinfo {author} {\bibfnamefont {K.}~\bibnamefont
  {Jacobs}},\ }\href {http://dx.doi.org/10.1038/s41534-020-00299-6} {\bibfield
  {journal} {\bibinfo  {journal} {npj Quantum Information}\ }\textbf {\bibinfo
  {volume} {6}} (\bibinfo {year} {2020})}\BibitemShut {NoStop}%
\bibitem [{\citenamefont {Kir\ifmmode~\check{s}\else \v{s}\fi{}anskas}\ \emph
  {et~al.}(2018)\citenamefont {Kir\ifmmode~\check{s}\else \v{s}\fi{}anskas},
  \citenamefont {Francki\'e},\ and\ \citenamefont {Wacker}}]{kirvsanskas2018}%
  \BibitemOpen
  \bibfield  {author} {\bibinfo {author} {\bibfnamefont {G.}~\bibnamefont
  {Kir\ifmmode~\check{s}\else \v{s}\fi{}anskas}}, \bibinfo {author}
  {\bibfnamefont {M.}~\bibnamefont {Francki\'e}}, \ and\ \bibinfo {author}
  {\bibfnamefont {A.}~\bibnamefont {Wacker}},\ }\href {\doibase
  10.1103/PhysRevB.97.035432} {\bibfield  {journal} {\bibinfo  {journal} {Phys.
  Rev. B}\ }\textbf {\bibinfo {volume} {97}},\ \bibinfo {pages} {035432}
  (\bibinfo {year} {2018})}\BibitemShut {NoStop}%
\bibitem [{\citenamefont {Nielsen}\ and\ \citenamefont {Chuang}(2011)}]{NnC}%
  \BibitemOpen
  \bibfield  {author} {\bibinfo {author} {\bibfnamefont {M.~A.}\ \bibnamefont
  {Nielsen}}\ and\ \bibinfo {author} {\bibfnamefont {I.~L.}\ \bibnamefont
  {Chuang}},\ }\href@noop {} {\emph {\bibinfo {title} {Quantum Computation and
  Quantum Information: 10th Anniversary Edition}}},\ \bibinfo {edition} {10th}\
  ed.\ (\bibinfo  {publisher} {Cambridge University Press},\ \bibinfo {address}
  {USA},\ \bibinfo {year} {2011})\BibitemShut {NoStop}%
\bibitem [{\citenamefont {Bennett}\ and\ \citenamefont
  {DiVincenzo}(2000)}]{Bennett2000}%
  \BibitemOpen
  \bibfield  {author} {\bibinfo {author} {\bibfnamefont {C.~H.}\ \bibnamefont
  {Bennett}}\ and\ \bibinfo {author} {\bibfnamefont {D.~P.}\ \bibnamefont
  {DiVincenzo}},\ }\href {\doibase 10.1038/35005001} {\bibfield  {journal}
  {\bibinfo  {journal} {Nature}\ }\textbf {\bibinfo {volume} {404}},\ \bibinfo
  {pages} {247} (\bibinfo {year} {2000})}\BibitemShut {NoStop}%
\bibitem [{\citenamefont {Streltsov}\ \emph {et~al.}(2017)\citenamefont
  {Streltsov}, \citenamefont {Adesso},\ and\ \citenamefont
  {Plenio}}]{Streltsov_2017}%
  \BibitemOpen
  \bibfield  {author} {\bibinfo {author} {\bibfnamefont {A.}~\bibnamefont
  {Streltsov}}, \bibinfo {author} {\bibfnamefont {G.}~\bibnamefont {Adesso}}, \
  and\ \bibinfo {author} {\bibfnamefont {M.~B.}\ \bibnamefont {Plenio}},\
  }\href {\doibase 10.1103/RevModPhys.89.041003} {\bibfield  {journal}
  {\bibinfo  {journal} {Rev. Mod. Phys.}\ }\textbf {\bibinfo {volume} {89}},\
  \bibinfo {pages} {041003} (\bibinfo {year} {2017})}\BibitemShut {NoStop}%
\bibitem [{\citenamefont {Lostaglio}\ \emph {et~al.}(2015)\citenamefont
  {Lostaglio}, \citenamefont {Korzekwa}, \citenamefont {Jennings},\ and\
  \citenamefont {Rudolph}}]{Lostaglio_2015}%
  \BibitemOpen
  \bibfield  {author} {\bibinfo {author} {\bibfnamefont {M.}~\bibnamefont
  {Lostaglio}}, \bibinfo {author} {\bibfnamefont {K.}~\bibnamefont {Korzekwa}},
  \bibinfo {author} {\bibfnamefont {D.}~\bibnamefont {Jennings}}, \ and\
  \bibinfo {author} {\bibfnamefont {T.}~\bibnamefont {Rudolph}},\ }\href
  {\doibase 10.1103/PhysRevX.5.021001} {\bibfield  {journal} {\bibinfo
  {journal} {Phys. Rev. X}\ }\textbf {\bibinfo {volume} {5}},\ \bibinfo {pages}
  {021001} (\bibinfo {year} {2015})}\BibitemShut {NoStop}%
\bibitem [{\citenamefont {Narasimhachar}\ and\ \citenamefont
  {Gour}(2015)}]{Narasimhachar_2015}%
  \BibitemOpen
  \bibfield  {author} {\bibinfo {author} {\bibfnamefont {V.}~\bibnamefont
  {Narasimhachar}}\ and\ \bibinfo {author} {\bibfnamefont {G.}~\bibnamefont
  {Gour}},\ }\href {http://dx.doi.org/10.1038/ncomms8689} {\bibfield  {journal}
  {\bibinfo  {journal} {Nature Communications}\ }\textbf {\bibinfo {volume}
  {6}} (\bibinfo {year} {2015})}\BibitemShut {NoStop}%
\bibitem [{\citenamefont {Mitchison}\ \emph {et~al.}(2015)\citenamefont
  {Mitchison}, \citenamefont {Woods}, \citenamefont {Prior},\ and\
  \citenamefont {Huber}}]{Mitchison_2015}%
  \BibitemOpen
  \bibfield  {author} {\bibinfo {author} {\bibfnamefont {M.~T.}\ \bibnamefont
  {Mitchison}}, \bibinfo {author} {\bibfnamefont {M.~P.}\ \bibnamefont
  {Woods}}, \bibinfo {author} {\bibfnamefont {J.}~\bibnamefont {Prior}}, \ and\
  \bibinfo {author} {\bibfnamefont {M.}~\bibnamefont {Huber}},\ }\href
  {\doibase 10.1088/1367-2630/17/11/115013} {\bibfield  {journal} {\bibinfo
  {journal} {New Journal of Physics}\ }\textbf {\bibinfo {volume} {17}},\
  \bibinfo {pages} {115013} (\bibinfo {year} {2015})}\BibitemShut {NoStop}%
\bibitem [{\citenamefont {Allahverdyan}\ \emph {et~al.}(2004)\citenamefont
  {Allahverdyan}, \citenamefont {Balian},\ and\ \citenamefont
  {Nieuwenhuizen}}]{Allahverdyan_2004}%
  \BibitemOpen
  \bibfield  {author} {\bibinfo {author} {\bibfnamefont {A.~E.}\ \bibnamefont
  {Allahverdyan}}, \bibinfo {author} {\bibfnamefont {R.}~\bibnamefont
  {Balian}}, \ and\ \bibinfo {author} {\bibfnamefont {T.~M.}\ \bibnamefont
  {Nieuwenhuizen}},\ }\href {http://dx.doi.org/10.1209/epl/i2004-10101-2}
  {\bibfield  {journal} {\bibinfo  {journal} {Europhysics Letters (EPL)}\
  }\textbf {\bibinfo {volume} {67}},\ \bibinfo {pages} {565–571} (\bibinfo
  {year} {2004})}\BibitemShut {NoStop}%
\bibitem [{\citenamefont {Korzekwa}\ \emph {et~al.}(2016)\citenamefont
  {Korzekwa}, \citenamefont {Lostaglio}, \citenamefont {Oppenheim},\ and\
  \citenamefont {Jennings}}]{Korzekwa_2016}%
  \BibitemOpen
  \bibfield  {author} {\bibinfo {author} {\bibfnamefont {K.}~\bibnamefont
  {Korzekwa}}, \bibinfo {author} {\bibfnamefont {M.}~\bibnamefont {Lostaglio}},
  \bibinfo {author} {\bibfnamefont {J.}~\bibnamefont {Oppenheim}}, \ and\
  \bibinfo {author} {\bibfnamefont {D.}~\bibnamefont {Jennings}},\ }\href
  {http://dx.doi.org/10.1088/1367-2630/18/2/023045} {\bibfield  {journal}
  {\bibinfo  {journal} {New Journal of Physics}\ }\textbf {\bibinfo {volume}
  {18}},\ \bibinfo {pages} {023045} (\bibinfo {year} {2016})}\BibitemShut
  {NoStop}%
\bibitem [{\citenamefont {Kammerlander}\ and\ \citenamefont
  {Anders}(2016)}]{Kammerlander_2016}%
  \BibitemOpen
  \bibfield  {author} {\bibinfo {author} {\bibfnamefont {P.}~\bibnamefont
  {Kammerlander}}\ and\ \bibinfo {author} {\bibfnamefont {J.}~\bibnamefont
  {Anders}},\ }\href {http://dx.doi.org/10.1038/srep22174} {\bibfield
  {journal} {\bibinfo  {journal} {Scientific Reports}\ }\textbf {\bibinfo
  {volume} {6}} (\bibinfo {year} {2016})}\BibitemShut {NoStop}%
\bibitem [{\citenamefont {Santos}\ \emph {et~al.}(2019)\citenamefont {Santos},
  \citenamefont {Céleri}, \citenamefont {Landi},\ and\ \citenamefont
  {Paternostro}}]{Santos_2019}%
  \BibitemOpen
  \bibfield  {author} {\bibinfo {author} {\bibfnamefont {J.~P.}\ \bibnamefont
  {Santos}}, \bibinfo {author} {\bibfnamefont {L.~C.}\ \bibnamefont {Céleri}},
  \bibinfo {author} {\bibfnamefont {G.~T.}\ \bibnamefont {Landi}}, \ and\
  \bibinfo {author} {\bibfnamefont {M.}~\bibnamefont {Paternostro}},\ }\href
  {http://dx.doi.org/10.1038/s41534-019-0138-y} {\bibfield  {journal} {\bibinfo
   {journal} {npj Quantum Information}\ }\textbf {\bibinfo {volume} {5}}
  (\bibinfo {year} {2019})}\BibitemShut {NoStop}%
\bibitem [{\citenamefont {Francica}\ \emph {et~al.}(2019)\citenamefont
  {Francica}, \citenamefont {Goold},\ and\ \citenamefont
  {Plastina}}]{Francica_2019}%
  \BibitemOpen
  \bibfield  {author} {\bibinfo {author} {\bibfnamefont {G.}~\bibnamefont
  {Francica}}, \bibinfo {author} {\bibfnamefont {J.}~\bibnamefont {Goold}}, \
  and\ \bibinfo {author} {\bibfnamefont {F.}~\bibnamefont {Plastina}},\ }\href
  {\doibase 10.1103/PhysRevE.99.042105} {\bibfield  {journal} {\bibinfo
  {journal} {Phys. Rev. E}\ }\textbf {\bibinfo {volume} {99}},\ \bibinfo
  {pages} {042105} (\bibinfo {year} {2019})}\BibitemShut {NoStop}%
\bibitem [{\citenamefont {Fleming}\ and\ \citenamefont
  {Cummings}(2011)}]{fleming_cummings_accuracy}%
  \BibitemOpen
  \bibfield  {author} {\bibinfo {author} {\bibfnamefont {C.~H.}\ \bibnamefont
  {Fleming}}\ and\ \bibinfo {author} {\bibfnamefont {N.~I.}\ \bibnamefont
  {Cummings}},\ }\href {\doibase 10.1103/PhysRevE.83.031117} {\bibfield
  {journal} {\bibinfo  {journal} {Phys. Rev. E}\ }\textbf {\bibinfo {volume}
  {83}},\ \bibinfo {pages} {031117} (\bibinfo {year} {2011})}\BibitemShut
  {NoStop}%
\bibitem [{\citenamefont {Purkayastha}\ \emph {et~al.}(2020)\citenamefont
  {Purkayastha}, \citenamefont {Guarnieri}, \citenamefont {Mitchison},
  \citenamefont {Filip},\ and\ \citenamefont {Goold}}]{Archak_2020}%
  \BibitemOpen
  \bibfield  {author} {\bibinfo {author} {\bibfnamefont {A.}~\bibnamefont
  {Purkayastha}}, \bibinfo {author} {\bibfnamefont {G.}~\bibnamefont
  {Guarnieri}}, \bibinfo {author} {\bibfnamefont {M.~T.}\ \bibnamefont
  {Mitchison}}, \bibinfo {author} {\bibfnamefont {R.}~\bibnamefont {Filip}}, \
  and\ \bibinfo {author} {\bibfnamefont {J.}~\bibnamefont {Goold}},\ }\href
  {http://dx.doi.org/10.1038/s41534-020-0256-6} {\bibfield  {journal} {\bibinfo
   {journal} {npj Quantum Information}\ }\textbf {\bibinfo {volume} {6}}
  (\bibinfo {year} {2020})}\BibitemShut {NoStop}%
\bibitem [{\citenamefont {Johansson}\ \emph {et~al.}(2012)\citenamefont
  {Johansson}, \citenamefont {Nation},\ and\ \citenamefont {Nori}}]{qutip_1}%
  \BibitemOpen
  \bibfield  {author} {\bibinfo {author} {\bibfnamefont {J.}~\bibnamefont
  {Johansson}}, \bibinfo {author} {\bibfnamefont {P.}~\bibnamefont {Nation}}, \
  and\ \bibinfo {author} {\bibfnamefont {F.}~\bibnamefont {Nori}},\ }\href
  {https://www.sciencedirect.com/science/article/pii/S0010465512000835}
  {\bibfield  {journal} {\bibinfo  {journal} {Computer Physics Communications}\
  }\textbf {\bibinfo {volume} {183}},\ \bibinfo {pages} {1760} (\bibinfo {year}
  {2012})}\BibitemShut {NoStop}%
\bibitem [{\citenamefont {Johansson}\ \emph {et~al.}(2013)\citenamefont
  {Johansson}, \citenamefont {Nation},\ and\ \citenamefont {Nori}}]{qutip_2}%
  \BibitemOpen
  \bibfield  {author} {\bibinfo {author} {\bibfnamefont {J.}~\bibnamefont
  {Johansson}}, \bibinfo {author} {\bibfnamefont {P.}~\bibnamefont {Nation}}, \
  and\ \bibinfo {author} {\bibfnamefont {F.}~\bibnamefont {Nori}},\ }\href
  {\doibase https://doi.org/10.1016/j.cpc.2012.11.019} {\bibfield  {journal}
  {\bibinfo  {journal} {Computer Physics Communications}\ }\textbf {\bibinfo
  {volume} {184}},\ \bibinfo {pages} {1234} (\bibinfo {year}
  {2013})}\BibitemShut {NoStop}%
\bibitem [{\citenamefont {Guarnieri}\ \emph {et~al.}(2018)\citenamefont
  {Guarnieri}, \citenamefont {Kol\'a\ifmmode~\check{r}\else \v{r}\fi{}},\ and\
  \citenamefont {Filip}}]{Guarnieri_2018}%
  \BibitemOpen
  \bibfield  {author} {\bibinfo {author} {\bibfnamefont {G.}~\bibnamefont
  {Guarnieri}}, \bibinfo {author} {\bibfnamefont {M.}~\bibnamefont
  {Kol\'a\ifmmode~\check{r}\else \v{r}\fi{}}}, \ and\ \bibinfo {author}
  {\bibfnamefont {R.}~\bibnamefont {Filip}},\ }\href {\doibase
  10.1103/PhysRevLett.121.070401} {\bibfield  {journal} {\bibinfo  {journal}
  {Phys. Rev. Lett.}\ }\textbf {\bibinfo {volume} {121}},\ \bibinfo {pages}
  {070401} (\bibinfo {year} {2018})}\BibitemShut {NoStop}%
\bibitem [{\citenamefont {Thingna}\ \emph {et~al.}(2012)\citenamefont
  {Thingna}, \citenamefont {Wang},\ and\ \citenamefont
  {H{\"a}nggi}}]{Juzar_2012}%
  \BibitemOpen
  \bibfield  {author} {\bibinfo {author} {\bibfnamefont {J.}~\bibnamefont
  {Thingna}}, \bibinfo {author} {\bibfnamefont {J.-S.}\ \bibnamefont {Wang}}, \
  and\ \bibinfo {author} {\bibfnamefont {P.}~\bibnamefont {H{\"a}nggi}},\
  }\href {\doibase 10.1063/1.4718706} {\bibfield  {journal} {\bibinfo
  {journal} {The Journal of Chemical Physics}\ }\textbf {\bibinfo {volume}
  {136}},\ \bibinfo {pages} {194110} (\bibinfo {year} {2012})}\BibitemShut
  {NoStop}%
\bibitem [{\citenamefont {Thingna}\ \emph {et~al.}(2013)\citenamefont
  {Thingna}, \citenamefont {Wang},\ and\ \citenamefont
  {H\"anggi}}]{Juzar_2013}%
  \BibitemOpen
  \bibfield  {author} {\bibinfo {author} {\bibfnamefont {J.}~\bibnamefont
  {Thingna}}, \bibinfo {author} {\bibfnamefont {J.-S.}\ \bibnamefont {Wang}}, \
  and\ \bibinfo {author} {\bibfnamefont {P.}~\bibnamefont {H\"anggi}},\ }\href
  {\doibase 10.1103/PhysRevE.88.052127} {\bibfield  {journal} {\bibinfo
  {journal} {Phys. Rev. E}\ }\textbf {\bibinfo {volume} {88}},\ \bibinfo
  {pages} {052127} (\bibinfo {year} {2013})}\BibitemShut {NoStop}%
\bibitem [{\citenamefont {Xu}\ \emph {et~al.}(2017)\citenamefont {Xu},
  \citenamefont {Thingna},\ and\ \citenamefont {Wang}}]{Juzar_2017}%
  \BibitemOpen
  \bibfield  {author} {\bibinfo {author} {\bibfnamefont {X.}~\bibnamefont
  {Xu}}, \bibinfo {author} {\bibfnamefont {J.}~\bibnamefont {Thingna}}, \ and\
  \bibinfo {author} {\bibfnamefont {J.-S.}\ \bibnamefont {Wang}},\ }\href
  {\doibase 10.1103/PhysRevB.95.035428} {\bibfield  {journal} {\bibinfo
  {journal} {Phys. Rev. B}\ }\textbf {\bibinfo {volume} {95}},\ \bibinfo
  {pages} {035428} (\bibinfo {year} {2017})}\BibitemShut {NoStop}%
\bibitem [{\citenamefont {Cattaneo}\ \emph {et~al.}(2021)\citenamefont
  {Cattaneo}, \citenamefont {De~Chiara}, \citenamefont {Maniscalco},
  \citenamefont {Zambrini},\ and\ \citenamefont {Giorgi}}]{collision1}%
  \BibitemOpen
  \bibfield  {author} {\bibinfo {author} {\bibfnamefont {M.}~\bibnamefont
  {Cattaneo}}, \bibinfo {author} {\bibfnamefont {G.}~\bibnamefont {De~Chiara}},
  \bibinfo {author} {\bibfnamefont {S.}~\bibnamefont {Maniscalco}}, \bibinfo
  {author} {\bibfnamefont {R.}~\bibnamefont {Zambrini}}, \ and\ \bibinfo
  {author} {\bibfnamefont {G.~L.}\ \bibnamefont {Giorgi}},\ }\href {\doibase
  10.1103/PhysRevLett.126.130403} {\bibfield  {journal} {\bibinfo  {journal}
  {Phys. Rev. Lett.}\ }\textbf {\bibinfo {volume} {126}},\ \bibinfo {pages}
  {130403} (\bibinfo {year} {2021})}\BibitemShut {NoStop}%
\bibitem [{\citenamefont {Brenes}\ \emph {et~al.}(2020)\citenamefont {Brenes},
  \citenamefont {Mendoza-Arenas}, \citenamefont {Purkayastha}, \citenamefont
  {Mitchison}, \citenamefont {Clark},\ and\ \citenamefont
  {Goold}}]{thermleads}%
  \BibitemOpen
  \bibfield  {author} {\bibinfo {author} {\bibfnamefont {M.}~\bibnamefont
  {Brenes}}, \bibinfo {author} {\bibfnamefont {J.~J.}\ \bibnamefont
  {Mendoza-Arenas}}, \bibinfo {author} {\bibfnamefont {A.}~\bibnamefont
  {Purkayastha}}, \bibinfo {author} {\bibfnamefont {M.~T.}\ \bibnamefont
  {Mitchison}}, \bibinfo {author} {\bibfnamefont {S.~R.}\ \bibnamefont
  {Clark}}, \ and\ \bibinfo {author} {\bibfnamefont {J.}~\bibnamefont
  {Goold}},\ }\href {\doibase 10.1103/PhysRevX.10.031040} {\bibfield  {journal}
  {\bibinfo  {journal} {Phys. Rev. X}\ }\textbf {\bibinfo {volume} {10}},\
  \bibinfo {pages} {031040} (\bibinfo {year} {2020})}\BibitemShut {NoStop}%
\end{thebibliography}%

\end{document}